%% file: conference_101719.tex
\documentclass[journal]{IEEEtran}
\IEEEoverridecommandlockouts
\usepackage{cite}
\usepackage{amsmath,amssymb,amsfonts}
\usepackage{algorithm}
\usepackage{algorithmicx}
\usepackage{algpseudocode}
\usepackage{graphicx}
\usepackage{textcomp}
\usepackage[table]{xcolor}
\usepackage{multirow}
\usepackage{array}
\usepackage{float}
\usepackage{url}
\usepackage[utf8]{inputenc}
\usepackage{textgreek}
\usepackage[T1]{fontenc}
\usepackage{pifont}
\usepackage{booktabs,threeparttable,makecell}
\usepackage[switch,mathlines]{lineno}
\usepackage{enumitem}
\usepackage{color,soul}
\usepackage{pifont}
\usepackage[normalem]{ulem}
\newcommand{\cmark}{\ding{51}}
\newcommand{\xmark}{\ding{55}}
\newcolumntype{P}[1]{>{\raggedright\arraybackslash}p{#1}}

\def\BibTeX{{\rm B\kern-.05em{\sc i\kern-.025em b}\kern-.08em
    T\kern-.1667em\lower.7ex\hbox{E}\kern-.125emX}}

\newcommand{\revv}[1]{\textcolor{black}{#1}}
\newcommand{\revvv}[1]{\textcolor{black}{#1}}
\newcommand{\rev}[1]{\textcolor{black}{#1}}

\begin{document}

\title{DistNN: Design-Space Exploration of Energy-Efficient Distributed Neural Networks in Low-Power Wearables\\
}

\author{\IEEEauthorblockN{1\textsuperscript{st} Given Name Surname}
\IEEEauthorblockA{\textit{dept. name of organization (of Aff.)} \\
\textit{name of organization (of Aff.)}\\
City, Country \\
email address or ORCID}
\and
\IEEEauthorblockN{2\textsuperscript{nd} Given Name Surname}
\IEEEauthorblockA{\textit{dept. name of organization (of Aff.)} \\
\textit{name of organization (of Aff.)}\\
City, Country \\
email address or ORCID}
\and
\IEEEauthorblockN{3\textsuperscript{rd} Given Name Surname}
\IEEEauthorblockA{\textit{dept. name of organization (of Aff.)} \\
\textit{name of organization (of Aff.)}\\
City, Country \\
email address or ORCID}
\and
\IEEEauthorblockN{4\textsuperscript{th} Given Name Surname}
\IEEEauthorblockA{\textit{dept. name of organization (of Aff.)} \\
\textit{name of organization (of Aff.)}\\
City, Country \\
email address or ORCID}
\and
\IEEEauthorblockN{5\textsuperscript{th} Given Name Surname}
\IEEEauthorblockA{\textit{dept. name of organization (of Aff.)} \\
\textit{name of organization (of Aff.)}\\
City, Country \\
email address or ORCID}
\and
\IEEEauthorblockN{6\textsuperscript{th} Given Name Surname}
\IEEEauthorblockA{\textit{dept. name of organization (of Aff.)} \\
\textit{name of organization (of Aff.)}\\
City, Country \\
email address or ORCID}
}

\author{
        Ming-Che Li*,~\IEEEmembership{Student~Member,~IEEE}, 
        Archisman Ghosh*,~\IEEEmembership{Student~Member,~IEEE},\\
        and~Shreyas~Sen,~\IEEEmembership{Senior~Member,~IEEE.} \\
        \newline *Authors contributed equally}

        \author{Meghna Roy Chowdhury,~\IEEEmembership{Student,~IEEE,} Ming-Che Li, ~\IEEEmembership{Student,~IEEE, }Archisman Ghosh, \\
        ~\IEEEmembership{Student,~IEEE,}  Md Faizul Bari,~\IEEEmembership{Student,~IEEE, }Shreyas Sen,~\IEEEmembership{Member,~IEEE.}}

\author{Meghna~Roy~Chowdhury,~\IEEEmembership{Student,~IEEE,}
        Ming\mbox{-}Che~Li*,~\IEEEmembership{Student,~IEEE,}
        Archisman~Ghosh*,~\IEEEmembership{Student,~IEEE,}
        Md~Faizul~Bari,~\IEEEmembership{Student,~IEEE,}
        and~Shreyas~Sen,~\IEEEmembership{Member,~IEEE}%
\thanks{The authors are with the Elmore Family School of Electrical and Computer Engineering, Purdue University, West Lafayette, Indiana, USA (e-mail: mroycho@purdue.edu; li4509@purdue.edu; ghosh69@purdue.edu; mbari@purdue.edu; shreyas@purdue.edu)}%
\thanks{* Li and Ghosh contributed equally to this work.}%
}

  \vspace{-5cm}      
\maketitle
 \begin{abstract}

Wearable devices are increasingly expected to support intelligent tasks such as augmented reality, continuous health monitoring, and vision-based interaction. However, the neural networks (NN) needed for these tasks often require memory and energy beyond the limits of wearable-class nodes. Cloud offloading reduces node-side computation, but it increases communication demand and may introduce latency and privacy concerns. TinyML reduces model complexity, but it can reduce accuracy and does not always resolve the memory limits of vision workloads. \revvv{This paper presents DistNN, a hardware-software framework for distributed neural network execution on wearables, in which split inference is used to partition a neural network between an ultra-low-power (ULP) wearable node and a nearby resource-rich hub (e.g. smartphone).} The node executes the early layers and transmits intermediate feature maps to the hub, which completes the remaining computation. 
To select the distribution point \revv{(split point)}, we define a platform-independent figure of merit (FoM) based on data volume (DV) and cumulative computation cost (CC). This captures the trade-off in distributed inference, where an earlier distribution point reduces CC but increases communication cost, while a later distribution point reduces DV but increases cumulative CC. We further formalize the conditions under which distribution provides a net node-side energy benefit, and show that the target autoencoder (AE) workloads are not feasible as monolithic on-node executions on representative wearable microcontrollers. \rev{To support ULP operation, we synthesize a custom node using low-precision fixed-point arithmetic. Post-synthesis estimates place DistNN within the ULP ($<10mW$) node-power target, with node-compute power averaging $\sim11\times$ lower than the recent ML ASIC baselines evaluated. Across convolution NNs and AEs, DistNN maintains task quality, achieving SSIM $\sim$ 0.90 for image reconstruction using low-precision node features and a resource-rich hub decoder.} This suggests that DistNN can support high quality energy-efficient wearable vision workloads.
~\vspace{-1mm}
\end{abstract}

\begin{IEEEkeywords}
Machine Vision, Split Computing, Split Inference, Collaborative Edge Inference, Distributed Neural Networks, Low Power, Autoencoders, Wearables
~\vspace{-3mm}
\end{IEEEkeywords}


\input{1_intro_new}
\input{2_relevant_topics}

\input{3_a_motivation}
\input{3_proposed_idea_new}



\input{4_eval_rev}

\input{6_conclusion}

\bibliographystyle{ieeetr}
\bibliography{ref}

\end{document}

%% file: 1_intro_new.tex
\section{Introduction}

\begin{figure}[t]
  \centering
  \includegraphics[width=\linewidth]{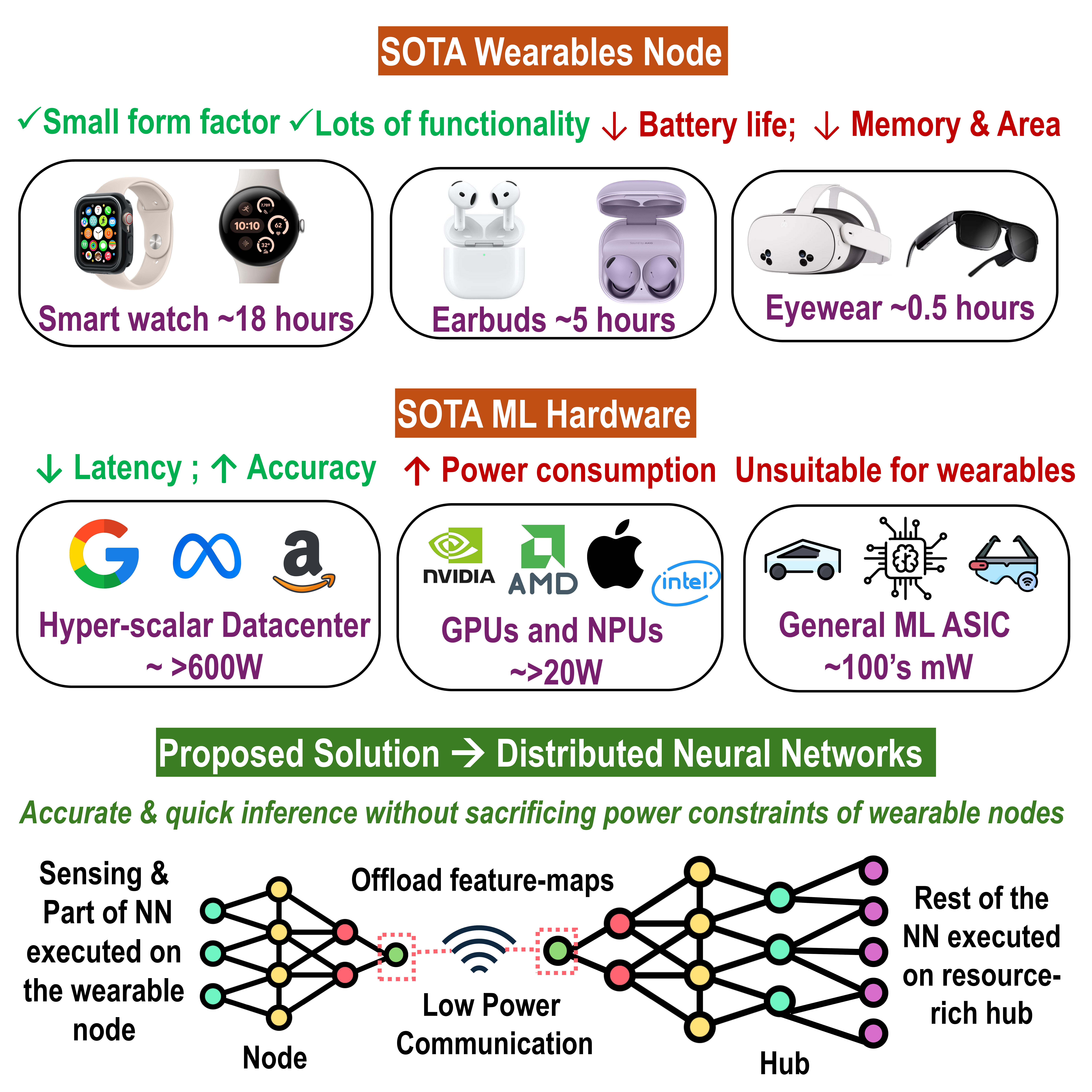}
  \vspace{-8mm}
   \begin{flushleft}
    {\footnotesize\emph{Logos and trademarks are the property of their respective owners and are used here for identification purposes only.}}
  \end{flushleft}
  \vspace{-5mm}
  \caption{Resource constraints in wearable ML systems. DistNN partitions NNs between an ultra-low-power wearable node and a nearby resource-rich hub to reduce the node's computation and power demands.}
    ~\vspace{-5mm}
  \label{fig_1_intro}
\end{figure}

\revvv{The rapid growth of the Internet of Things (IoT) and the Internet of Bodies (IoB) has driven the development of wearable devices that combine low-power sensors with communication technologies such as Bluetooth Low Energy (BLE), Zigbee, and near-field communication (NFC)~\cite{rawat2023harnessing,sen2024human}. Machine learning (ML) further expands their capabilities by supporting continuous health monitoring, activity recognition, and computer vision tasks such as image reconstruction, denoising, and segmentation. Running these workloads on wearable nodes is challenging because battery life, memory, and power are tightly constrained, whereas general-purpose ML hardware is designed for much larger resource budgets. As shown in Fig.~\ref{fig_1_intro}, commercial smartwatches last approximately 18 hours under mixed use, AirPods provide 4--5 hours of operation, and smart glasses sustain only 30--40 minutes of continuous streaming~\cite{AirPods421:online,RayBanMe3:online}. In comparison, GPUs commonly used for ML inference can draw hundreds of watts~\cite{ometov2021survey}. Table~\ref{tab:mlmodels} summarizes ML models and their architectural complexity.}

\begingroup
\begin{table}[t]
\centering
\caption{\label{tab:mlmodels}
Representative ML models \& their architectural complexity.}
\vspace{-2mm}

\footnotesize
\setlength{\tabcolsep}{3pt}
\renewcommand{\arraystretch}{1.05}

\resizebox{\columnwidth}{!}{%
\begin{tabular}{@{}llllr@{}}
\hline
\textbf{Model Family}
& \textbf{Application}
& \textbf{Model}
& \textbf{Layers}
& \textbf{Parameters} \\ \hline

\multirow{4}{*}{CNN}
& \multirow{2}{*}{Image classification}
& ResNet-50
& 50
& 25M \\
&
&
VGG-16
& 16
& 138M \\ \cline{2-5}

&
\multirow{2}{*}{Object detection}
& YOLOv3
& 106
& 62M \\
&
&
MobileNetV2
& 53
& 3.5M \\ \hline

\multirow{3}{*}{AE}
& Anomaly detection
& Denoising AE
& 20
& 1--2M \\

&
Image denoising
& U-Net
& 23
& 31M \\

&
Dimensionality reduction
& Variational AE
& 20
& 1--2M \\ \hline

\multirow{3}{*}{GAN}
& Super resolution
& SR-GAN
& 21
& 2M \\

&
Image-to-image translation
& Pix2Pix
& 15--20
& 11.5M \\

&
Image synthesis
& BigGAN
& 256
& 112M \\ \hline

\multirow{3}{*}{Transformer}
& Image classification
& ViT-B
& 12
& 86M \\

&
Object detection
& DETR
& 24
& 41M \\

&
Image segmentation
& Swin Transformer
& 12
& 60M \\ \hline
\end{tabular}%
}

\vspace{-4mm}
\end{table}
\endgroup

\revvv{Two approaches are commonly used to run these workloads under wearable resource limits. TinyML reduces model size through techniques such as quantization and pruning. However, they can lower accuracy and may not sufficiently reduce peak activation memory. A model may fit in storage yet still require too much memory to run. Cloud offloading instead sends raw sensor data to remote servers for processing. This reduces local computation and memory demands but introduces communication energy, latency, bandwidth use, and privacy risks. Neither approach jointly addresses on-node resource limits and the costs of cloud offloading. }



\revvv{DistNN addresses this gap by distributing neural-network (NN) execution between an ultra-low-power (ULP) wearable node and a nearby resource-rich hub. \footnote{\revvv{\textbf{Terminology.} We use \emph{distributed computing} for the overall node--hub cooperation. Partitioning NN execution between the node and hub is commonly known as \emph{split computing} or \emph{split inference}. Prior work refers to the boundary between the two partitions as a \emph{split point} or \emph{partition point}~\cite{matsubara2022split,parthasarathy2022partitioning}. In this paper, we call this boundary the \emph{distribution point}, consistent with DistNN and our preliminary work~\cite{chowdhury2024leveraging}. This difference is terminological, not technical. Source-specific terms are retained when discussing prior work.}} The node runs the early layers, transmits intermediate feature map to the hub, which completes the remaining layers. This reduces the node's computation while avoiding remote-cloud processing.}

\revvv{Existing split-inference methods reduce computation on resource-constrained nodes by partitioning neural networks, but most focus on classification and select the distribution point using platform-specific latency or energy models~\cite{bolat2025decentralized,gordon2012comet,li2022dissec,parthasarathy2022partitioning}. DistNN differs in two ways. It supports AE-based reconstruction and denoising in addition to classification, and it selects the distribution point before the target hardware platform or communication modality is chosen. Instead of requiring measurements from the target platform, DistNN uses the model architecture and tensor dimensions to rank candidate distribution points. These differences are important for deploying AEs on wearable devices. (\textbf{i)} At all quantization levels that preserve reconstruction quality, the evaluated AE models exceed the memory capacity of wearable-class MCUs. For example, the Nordic nRF52840 provides 1~MB of flash and 256~KB of SRAM, while the peak activation footprint alone exceeds the available SRAM by more than an order of magnitude (Section~\ref{sec:rq3_system}). For these models and this target platform, distribution is therefore a deployment requirement rather than an optimization. \textbf{(ii)} In reconstruction and denoising AEs, decoder-side feature-map expansion makes communication cost highly sensitive to the distribution point. Criteria based on classification-style feature compression can therefore select an unsuitable distribution point for these workloads. In addition, meeting the milliwatt-class ($\leq 10,\mathrm{mW}$) node-power target requires custom hardware rather than the commercial microcontrollers and embedded-vision platforms used in prior split-inference systems.}


The main contributions of this work are:

\begin{figure*}[t]
    \centering
    \includegraphics[width=0.8\linewidth]{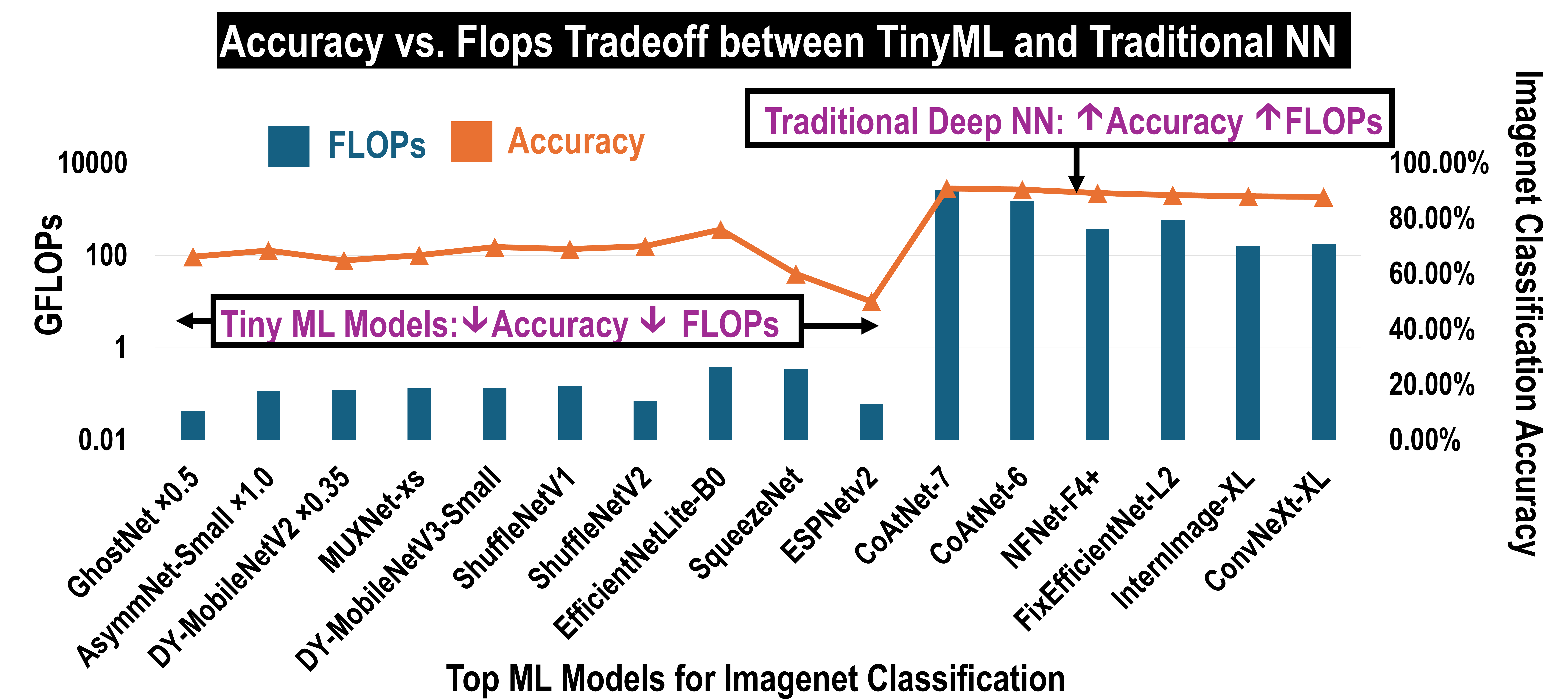}
    ~\vspace{-3mm}
    \caption{Trade-off between computational efficiency (FLOPs) and accuracy for DNN and TinyML models in ImageNet classification.}
    \vspace{-4mm}
    \label{fig_accVSflops}
\end{figure*}

\begin{enumerate}
    \item \revv{\textbf{Platform-independent distribution-point selection.} We introduce a figure of merit (FoM) that ranks distribution points by the normalized product of data volume and cumulative computation cost, from network topology alone, extending to energy \& latency aware forms once hardware-specific parameters are known. Unlike prior methods, which target mostly classification CNNs and apply only once the hardware and communication modality are fixed, the FoM covers both CNNs and AEs with no platform commitment.}

    \item \revv{\textbf{Node-side energy conditions for distributed execution.} We formalize the conditions under which distribution reduces node energy, including communication, \& validate them across low-power communication modalities.}


    \item \revv{\textbf{Distribution as a deployment requirement.} We show that the full AE models exceed the SRAM of common wearable-class MCUs even at low precision, whereas distribution allows them to fit and run on the node, establishing distribution as a requirement rather than an optimization. Furthermore, a custom low-precision (8-bit feature, 10-bit weight) fixed-point node, paired with a hub decoder, recovers reconstruction quality from SSIM 0.68 at the node to 0.90 after hub decoding, exploiting the asymmetry of DistNN.}
 
    \item \revv{\textbf{Ultra-low-power node through hardware-software co-design.} With our synthesized custom-MAC-based node, DistNN estimated node-compute power of $0.204$\,mW for AE1 \& $6.5$\,mW for AE2, averaging about $11\times$ lower than the recent ML ASIC baselines evaluated, with estimated end-to-end AE1 latency of $6.8$\,ms over Wi-R and $13.9$\,ms over BLE.}   
\end{enumerate}

\revv{Contributions~1 and~2 are methodological, whereas Contributions~3 and~4 present the implementation and experimental results.}
Together, these results show that model-guided split inference, low-precision node hardware, and hub-side decoding can support AE-based vision workloads that exceed the memory and power limits of monolithic execution on wearable-class microcontrollers.
~\vspace{-3mm}

%% file: 2_relevant_topics.tex
\section{Relevant Topics}
\label{sec:lit}
The challenges outlined above show that while wearable and IoB devices have immense potential, they remain constrained by energy and computational limitations. To better situate our approach within this landscape, we review existing research in TinyML, data offloading, and hardware acceleration, which together frame the design space our work builds upon. \vspace{-4mm}


\subsection{TinyML and its accuracy challenges}

TinyML has gained traction since it emerged as a way to run machine learning on resource-constrained devices for wearable computing, where low computational overhead and cost are decisive~\cite{Abadade2023A,Tsoukas2021A,Schizas2022TinyML}. By keeping inference on the device, it also preserves privacy and avoids sending sensitive data to external servers~\cite{Sanchez-Iborra2020TinyML-Enabled}. To meet the tight compute, memory, and energy budgets of IoT devices, frameworks such as TensorFlow Lite apply quantization, pruning, and model distillation to traditional DNNs. Despite these optimizations, preserving accuracy under such constraints remains a pressing challenge~\cite{lin2023tiny}. As shown in Fig.~\ref{fig_accVSflops}, DNNs reach far higher ImageNet accuracy but demand orders of magnitude more FLOPs, whereas TinyML trades accuracy for efficiency (plot from~\cite{ImageNet88:online}). Scalability compounds the problem, since heterogeneous IoT hardware demands robust device management and standardized protocols for consistent performance~\cite{Szydlo2023Device}. In all, TinyML brings learning on-device with stronger privacy, but its limits in accuracy, scalability, and compute, especially for computer vision, motivate offloading intensive computation from wearables to resource-rich platforms. \vspace{-4mm}
 \subsection{Data offloading}
\begin{figure}[t]
    \centering
    \includegraphics[width=1\linewidth] {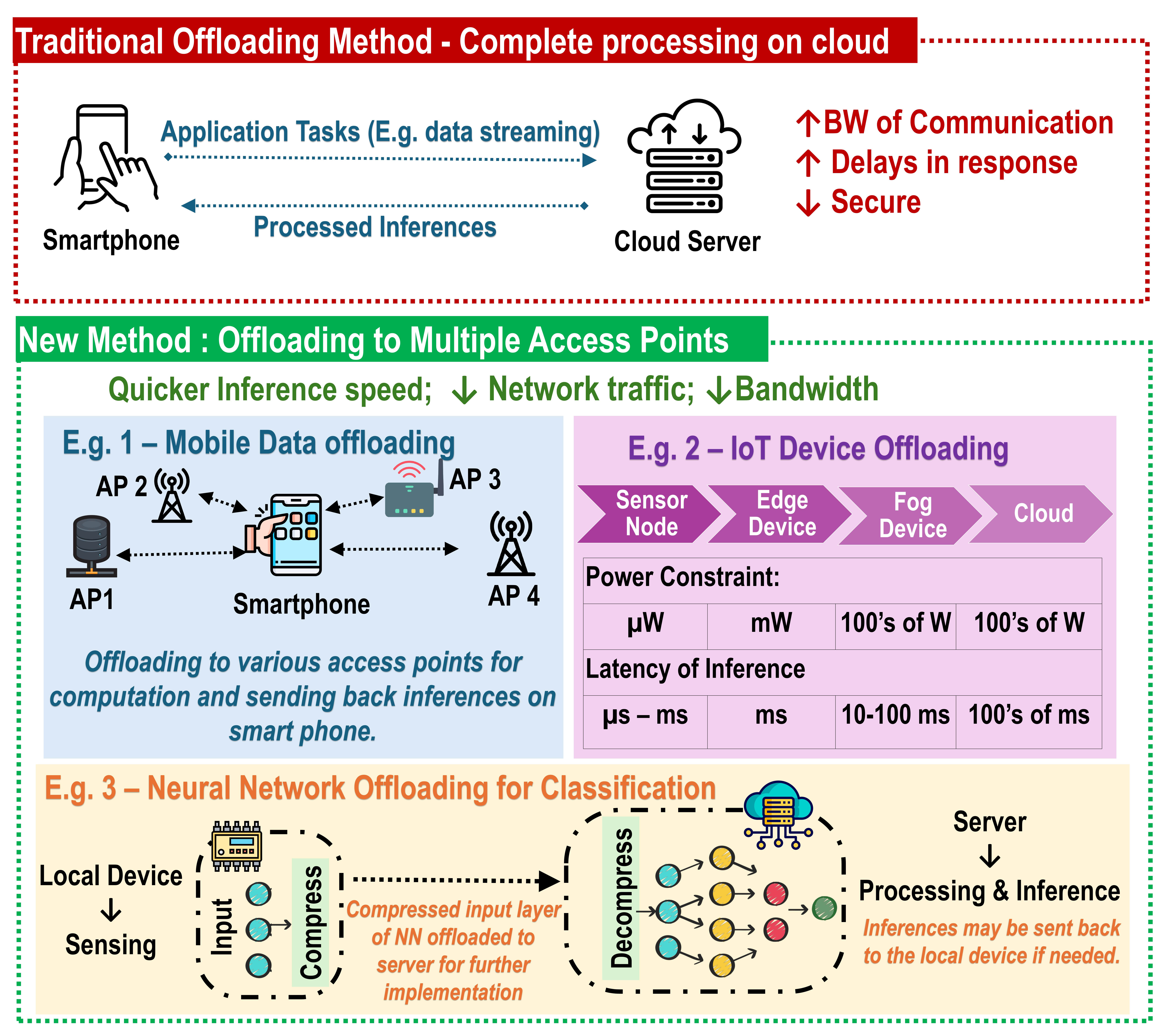}
      ~\vspace{-7mm}
    \caption{Comparison of traditional cloud-based processing with an offloading method to multiple access points. \vspace{-4mm}}
 \label{fig_do}
\end{figure}

 Data offloading transfers tasks from local devices to external resources such as cloud servers or edge devices, and is categorized into computation, storage, and communication offloading~\cite{Kumar2013A}. As depicted in Fig.~\ref{fig_do}, mobile data offloading, prevalent in video streaming, traditionally relies on cloud processing, incurring high bandwidth, cost, and delay. Newer approaches instead offload to nearby access points, speeding inference while reducing network traffic and device load~\cite{Park2019Video}. In IoT applications, offloading is similarly distributed across local, edge, fog, and cloud layers~\cite{Almutairi2021A,Wu2021EEDTO:}. Safety-critical tasks stay local for minimal latency, the edge and fog handle navigation and low-latency ADAS/V2V/V2I interactions, and the cloud absorbs heavy workloads such as over-the-air updates, model training, and predictive maintenance, optimizing resource use across the system~\cite{Zhou2018A}.


Beyond offloading whole tasks, a related line of work splits the neural network itself between device and server. This is an early form of the distributed inference this paper builds on, particularly for classification. Yang et al. share low-level feature maps between a camera and a gateway~\cite{yang2017distributed}; Pinkham et al. add on-sensor and on-device caches to accelerate ResNet~\cite{pinkham2021near}; Hu et al.'s EdgeFlow assigns execution units to edge devices to cut latency and traffic~\cite{hu2022distributed}; DeepCOD pairs a lightweight wearable autoencoder with a server-side GAN decoder for image classification~\cite{yao2020deep}; and Jeong et al.'s PerDNN dynamically partitions computation between clients and edge servers to reduce latency and cold starts~\cite{jeong2020perdnn}.


Together, these approaches show that offloading can relieve local devices of heavy computation, but they also expose persistent bottlenecks in latency, bandwidth, and privacy that software-level partitioning alone cannot resolve. This has motivated a parallel line of work on CNN hardware accelerators that address computational demand directly at the device.~\vspace{-4mm}

\begin{figure*}[t]
    \centering
    \includegraphics[width=0.9\linewidth]{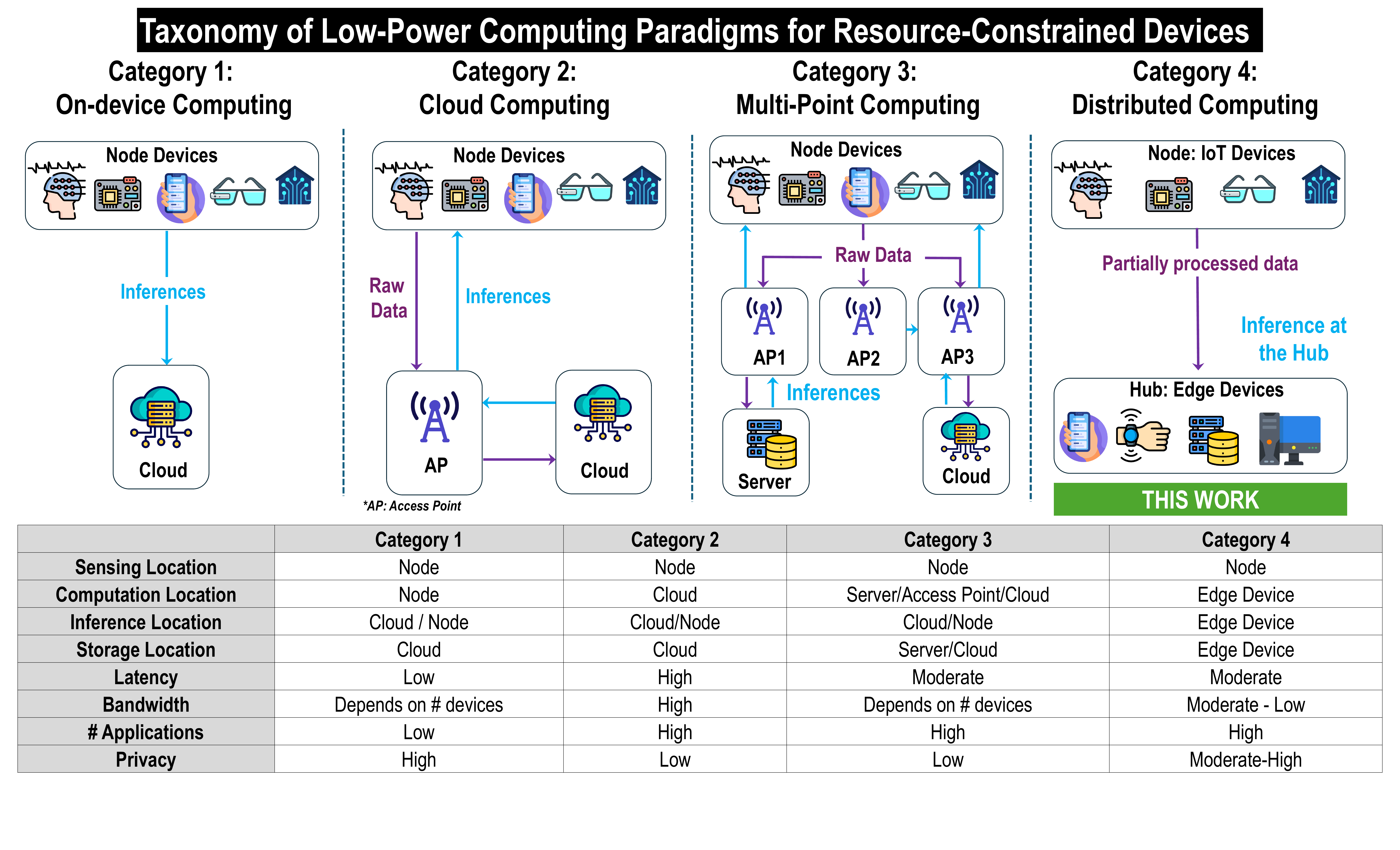}
    \vspace{-10mm}
    \caption{Taxonomy of low-power computing paradigms for resource-constrained systems. We compare \emph{on-device}, \emph{cloud}, \emph{multi-point}, and \emph{distributed} paradigms along sensing/computation/inference/storage location, latency, bandwidth, application scope, and privacy. This work emphasizes \emph{Distributed} (Category~4), where node devices transmit intermediate feature maps to nearby user devices, reducing energy and bandwidth while keeping data local.}
    \vspace{-4mm}
    \label{fig_lit}
    
\end{figure*}

\begin{table}[t]
\centering
\caption{State-of-the-art CNN hardware and performance.}
\label{fig_cnn_hw}
\vspace{-2mm}

\footnotesize
\setlength{\tabcolsep}{3pt}
\renewcommand{\arraystretch}{1.05}

\begin{tabular}{@{}p{1.5cm}p{3.2cm}p{2.6cm}@{}}
\hline
\textbf{Category} & \textbf{Examples} & \textbf{Performance} \\ \hline

Edge GPU &
NVIDIA Jetson Nano, Jetson Xavier NX, Intel Movidius Myriad X &
$>500$ GOPS; 2--20\,W \\ \hline

Tensor Processing Unit &
Google Edge TPU, TPU v4, Coral USB Accelerator &
$>5$ TOPS; $\sim$200\,W \\ \hline

Graphics Processing Unit &
NVIDIA A100, H100, RTX 3090, AMD MI250 &
$>50$ TOPS (mostly FP32); $>400$\,W \\ \hline

Neural Processing Unit &
Apple M4 (3\,nm), Qualcomm Hexagon NPU, Samsung Exynos NPU &
5--40 TOPS; lower power \\ \hline

Specialized AI Processor &
Cerebras CS-2, Amazon Inferentia, Tesla Dojo &
200 TOPS--100 ExaFLOPS; high power \\ \hline

Infrastructure Processing Unit &
Graphcore MK2 GC200 IPU (7\,nm), Graphcore Bow Pod &
250--350 TOPS; $\sim$300\,W \\ \hline
\end{tabular}

\vspace{-3mm}
\end{table}


\subsection{CNN Hardware}
Machine learning hardware spans data-center GPUs to specialized accelerators, but few platforms are built for the strict power and memory limits of wearables. Table~\ref{fig_cnn_hw} summarizes representative devices and their power characteristics. General-purpose GPUs such as NVIDIA's A100 deliver hundreds of TFLOPS at 300--400~W~\cite{NVIDIAA12:online}, and TPU v4 improves efficiency to 1.62~TOPS/W at 170~W~\cite{jouppi2023tpu}, a clear progress, yet still orders of magnitude above the milliwatt budgets of on-body inference. Even edge-oriented processors like Intel's Lunar Lake, with a 48~TOPS NPU and reduced SoC power~\cite{IntelsLu9:online}, do not reach $\mu W$ level for continuous use. FPGAs help in specific workloads. For instance, Stratix~IV implementations report 2.6$\times$–8$\times$ energy savings at minimal accuracy loss~\cite{das2024towards}, but limited on-chip memory, scalability, and sustained throughput constrain real-time wearable use.


In summary, current ML hardware targets general-purpose or edge-class workloads at relatively high power and precision, whereas wearables operate at milliwatt budgets and benefit from reduced-precision compute. This gap motivates dedicated ULP accelerators that sustain real-time inference within stringent energy and form-factor limits. \textit{\textbf{This paper explores the design space of DistNN and builds an optimized hardware design for the wearable node to perform low-power ML computations.}}
~\vspace{-3mm}

%% file: 3_a_motivation.tex
\section{Motivation}
\label{sec:motivation}

\begin{table*}[t]
\centering
\caption{\rev{Relevant literature across paradigms.}}
\vspace{-2mm}
\label{tab:lit_survey_b}
\setlength{\tabcolsep}{3pt}
\renewcommand{\arraystretch}{1.25}
\footnotesize
\begin{tabular}{P{0.9cm} P{1.3cm} P{1.5cm} P{1.2cm} P{0.5cm} P{3.5cm} P{1.8cm} P{3cm} P{2.8cm}}
\toprule
\textbf{Ref.} & \textbf{NN Type} & \textbf{Distribution (src $\rightarrow$ dst)} & \textbf{Workload Class} & \textbf{HW Impl.} & \textbf{Split-selection (where / what it minimizes)} & \textbf{Node Platform / Tech Node} & \textbf{Node Power} & \textbf{Latency} \\
\midrule


\rowcolor{cyan!20}

 \textbf{This work} & \textbf{Classif. + AE} & \textbf{Wearable MCU $\rightarrow$ Mobile} & \textbf{AE (expanding}) & \textbf{\checkmark} & \textbf{Splits at FoM-min layer / minimizes DV $\times$ cum. MACs (arch.-only)} & \textbf{Custom RTL, synthesized in 65\,nm} & \textbf{AE1: 0.208\,mW (Wi-R), 0.788\,mW (BLE); AE2: 8.43\,mW(Wi-R) (compute + comm.)} & \textbf{6.8\,ms (Wi-R), 13.9\,ms (BLE) per frame;} \\


\rowcolor{cyan!20}
\cite{bolat2025decentralized} & Classif.\ (LeNet) & MCU $\rightarrow$ MCU & Classif. & \xmark & Manual split at pool/flatten layers / minimizes transmitted data s.t.\ memory bound & ESP32-S3 (COTS, 40\,nm) & 170--377\,mW avg (incl.\ Wi-Fi); $\sim$508\,mW peak & 338.7\,ms E2E; 102.45\,ms/chunk \\

\rowcolor{cyan!20}
\cite{parthasarathy2022partitioning} & Classif. & Edge $\rightarrow$ Edge & Classif. & \xmark & 3-stage path search (longest/shortest/k-path) / minimizes bottleneck latency (sim.) & Simulated (synthetic net.\ model) & Not Reported & Rel.\ only ($\sim$10$\times$ vs.\ random, sim.) \\

\rowcolor{cyan!20}
\cite{li2022dissec} & CNN (conv-heavy) & Edge $\rightarrow$ Edge & Classif. & \xmark & Vertical conv.\ partitioning + A$^\ast$ search / minimizes global latency s.t.\ resource constraints & RPi 3B cluster ($\leq$5 dev., COTS) & Not Reported & 10--25\,ms (abs, config-dep.) \\

\rowcolor{cyan!20}
\cite{jayarajah2022comai} & Classif.\ (obj.\ det.) & Edge $\rightarrow$ Edge (P2P) & Classif. & \xmark & Runtime confidence-sharing at fixed layer / minimizes comm.\ (no split search) & Jetson Nano ($\sim$5\,W class) & 9160\,mW (full chain, incl.\ comms) & 536.8\,ms \\

\midrule

\rowcolor{yellow!20}
\cite{das2024towards} & Classif.\ (MVCNN) & Edge $\rightarrow$ Cloud & Multi-view 3D & \cmark & Fixed partitioning maximizing accuracy-over-energy (FLOPs + DV); energy from approx.\ computing & FPGA (Stratix IV, 40\,nm) + Nios II & Rel.\ only (2.6--8$\times$ E-savings) & Rel.\ (1.5--22.6$\times$ speedup) \\

\rowcolor{yellow!20}
\cite{laskaridis2020spinn} & CNN + early-exit & Mobile $\rightarrow$ Edge/Cloud & Classif. & \xmark & Co-optimizes split point + early-exit threshold / multi-objective (latency, accuracy, cost) & Jetson Xavier AGX & Rel.\ only & Rel.\ (83\% vs.\ Neurosurgeon) \\

\rowcolor{yellow!20}
\cite{hu2019dynamic} & Classif. + DAG nets & Edge $\rightarrow$ Cloud & Video analytics & \xmark & Min-cut (light load) / approx.\ search (heavy) / minimizes latency or throughput (load-adaptive) & RPi 3B (COTS) & Not Reported & Rel.\ (1.66--8.08$\times$); 0.59\,s @ 4\,Mbps \\

\rowcolor{yellow!20}
\cite{eshratifar2019jointdnn} & Discrim.\ + gen.\ + AE & Mobile $\leftrightarrow$ Cloud & Classif.\ / gen.\ / AE & \xmark & DAG shortest-path + ILP / minimizes latency or energy (selectable) & Jetson TX2 (Pascal) & Rel.\ only & Rel.\ only ($\sim$3--4$\times$) \\

\rowcolor{yellow!20}
\cite{kang2017neurosurgeon} & Classif. + speech/NLP & Mobile $\rightarrow$ Cloud & Classif.\ etc. & \xmark & Per-layer predicted cost model (platform-specific) / minimizes latency or energy & Jetson TK1 (Tegra K1) & Rel.\ only & Rel.\ (3.1$\times$ avg) \\

\midrule

\rowcolor{blue!20}
\cite{gordon2012comet} & Non--NN & CPU $\rightarrow$ Server & N/A & \xmark & $\tau$-scheduler thread migration (heuristic) / minimizes sync time & Smartphone (Captivate) & Not Reported (app-level only) & Rel.\ (2.88$\times$ WiFi, interactive) \\

\rowcolor{blue!20}
\cite{ra2011odessa} & Non--NN (classical CV) & Mobile $\rightarrow$ Server & N/A & \xmark & Stage-level greedy bottleneck adaptation / minimizes makespan + throughput & Netbook (Atom N270) & Not Reported & 0.5--9.2\,s (abs, app-dep.) \\

\bottomrule
\end{tabular}
\vspace{1.5mm}

\begin{tabular}{P{4.1cm} P{6.2cm} P{6.2cm}}
\toprule
\cellcolor{cyan!20}\centering\textbf{Category 4} &
\cellcolor{yellow!20}\centering\textbf{Cats.~2--3 with NN distribution(split inference)} &
\cellcolor{blue!20}\centering\textbf{Cats.~2--3 with non--NN offloading} \\

\end{tabular}

\vspace{1mm}
{\footnotesize 
The HW Impl.\ column marks \cmark\ for entries evaluated using custom hardware through RTL synthesis, custom silicon, or an FPGA prototype; \xmark\ denotes evaluation on commodity hardware or simulation. ``Rel.'' denotes a relative improvement reported without absolute node power; ``sim.'' denotes results from a simulated network model. Reported node-power figures for prior Category~4 work include wireless communication. Category~2--3 entries use mobile/cloud-class platforms where node-side power is not the operative metric. Memory footprint is discussed in Section~\ref{sec:rq3_system}. Power and platform figures are \emph{not} normalized to a common technology node.}
~\vspace{-5mm}
\end{table*}


As discussed in Section~\ref{sec:lit}, TinyML, data offloading, and specialized CNN hardware each address part of the accuracy, energy trade-off but none achieves ULP ($\leq$10mW) operation for wearable vision. To examine this gap systematically, we organize the design space into a taxonomy that clarifies where existing approaches reach their practical limits, and then identify the specific shortcomings that motivate DistNN. We organize resource-constrained machine-learning systems into four paradigms.
As illustrated in Fig.~\ref{fig_lit}, the taxonomy compares these paradigms across sensing, computation, inference, and storage location, together with latency, bandwidth, application scope, and privacy. This organization provides a structured view of the design space and serves as the basis for identifying the gaps relevant to wearable systems.

\subsubsection{\textbf{Category 1: On-Device Computing}}
On-device computing performs both sensing and inference locally on the node using TinyML models optimized through quantization and pruning. This minimizes latency and preserves privacy but is bounded by SRAM, compute throughput, and energy. Examples include smartwatch-based atrial fibrillation detection and EMG/EEG classification for fatigue or stress recognition, where only event flags or summary metrics are transmitted~\cite{bhamare2024tinyml,liu2019emg}. Such systems are energy-efficient but cannot scale to vision workloads with higher data dimensionality.

\subsubsection{\textbf{Category 2: Cloud Computing}}
Cloud computing offloads raw sensor data to remote servers for full-model inference. It removes node-side compute cost and enables complex analytics, but at the expense of latency, bandwidth, and privacy. This paradigm suits non-real-time or retrospective analytics, such as continuous glucose or ECG monitoring, where data are uploaded for trend extraction and diagnostic reporting. Its dependence on connectivity and its privacy exposure limit its suitability for wearable systems.

\subsubsection{\textbf{Category 3: Multi-Point Computing}}
Multi-point computing introduces intermediate fog or edge servers that process data closer to the source. This hierarchy reduces round-trip latency and bandwidth relative to cloud-only offloading while preserving inference accuracy. However, raw or minimally processed data typically still traverse shared networks, presenting privacy risks, and the architecture depends on available resources. It is effective in semi-centralized settings such as hospital gateways or factory edge nodes, but remains impractical for ad hoc wearable deployments.

\subsubsection{\textbf{Category 4: Distributed Computing}}
Distributed computing divides a neural network between a constrained node and a nearby resource-rich peer, such as a smartphone or laptop. The node executes the early layers to produce a compact feature map, which is transmitted to the peer to complete inference. This avoids the cloud, reduces transmitted data volume, and keeps data within a user-controlled local network. By exploiting the regular charging cycles of companion devices, it is an attractive model for next-generation wearable intelligence, and it is the paradigm this work targets. \revv{In the broader literature, this two-party (device–peer) partitioning is termed \emph{split computing} or \emph{split inference}~\cite{matsubara2022split}; Category~4 specializes it to the ULP wearable regime, distinguished from the mobile/edge-tier instances in Categories~2--3.} 

 \paragraph*{\textbf{\rev{Gaps in Existing Category~4 Approaches}  }}
\label{sec:gaps}
As summarized in Table~\ref{tab:lit_survey_b}, most split-computing research sits in Categories 2–3, partitioning networks across mobile–cloud or edge tiers with varied distribution-point selection logic. Neurosurgeon fits a per-layer cost model to the target device~\cite{kang2017neurosurgeon}; JointDNN solves partitioning as a shortest-path or integer linear program over a layer graph, covering generative and autoencoder as well as classification workloads~\cite{eshratifar2019jointdnn}; SPINN co-optimizes the split point with an early-exit threshold~\cite{laskaridis2020spinn}; and DRAX fixes a pooling point by maximizing accuracy-over-energy from edge FLOPs and output-feature-map size~\cite{das2024towards}. Yet all select the distribution point from \emph{platform-specific} latency or energy measured on the target hardware, on mobile, edge, or FPGA-class nodes drawing 100\,mW to several watts, not the ULP ($\leq$10\,mW) regime targeted in this work. A smaller, more recent body of Category~4 work pushes distribution onto genuinely constrained nodes. DISSEC cuts latency on Raspberry~Pi clusters via vertical convolutional partitioning~\cite{li2022dissec}; Parthasarathy et~al. minimize bottleneck latency on a simulated edge network~\cite{parthasarathy2022partitioning} and DDSNN distributes a LeNet classifier across ESP32-S3 microcontrollers without compression~\cite{bolat2025decentralized}. These establish feasibility but share two limitations that matter for wearable vision. First, they target \emph{classification} networks whose feature maps generally shrink toward the output, making DV alone a reasonable criterion for selecting the distribution point. Reconstruction and denoising autoencoders instead \emph{expand} data through the decoder, placing the most compressed layer in the network interior, so distribution-point selection must weigh DV against the computation already performed on the node, not merely the minimum feature size. Second, node-side power exceeds wearable budgets. DDSNN measures $170$--$377\,\mathrm{mW}$ per node on the ESP32-S3 with active Wi-Fi~\cite{bolat2025decentralized} and ComAI reports $9.1\,\mathrm{W}$ per node including communication~\cite{jayarajah2022comai}, while DISSEC~\cite{li2022dissec} and Parthasarathy et~al.~\cite{parthasarathy2022partitioning} report none, and most come from $40\,\mathrm{nm}$-and-above COTS parts not normalized to a common technology node. Wearable vision compounds microcontrollers such as the Nordic nRF52840 (Arduino Nano~33 BLE Sense with 1\,MB Flash, 256\,KB SRAM) which cannot fit an AE even under INT8 quantization, as the peak activation tensor still exceeds available SRAM (Section~\ref{sec:rq3_system}), making distribution \emph{necessary} rather than just advantageous. 
Operating within the ULP node-power target further requires coordinated design across distribution-point placement, feature-map representation, communication modality, and silicon substrate, a co-design space that the commodity-MCU and embedded-vision platforms of prior Category~4 work leave unaddressed. \textbf{This work extends Category~4 distributed computing to the ULP wearable regime through DistNN.} \revv{In this work, the point at which the network is split or partitioned across devices is referred to as the \emph{distribution point}.} DistNN combines a platform-independent metric that balances data volume against cumulative node computation. The next section details the DistNN framework and methodology.

%% file: 3_proposed_idea_new.tex
\rev{ \section{Proposed DistNN Framework}}

\begin{figure*}[t]
    \centering
    \includegraphics [width=1\linewidth] {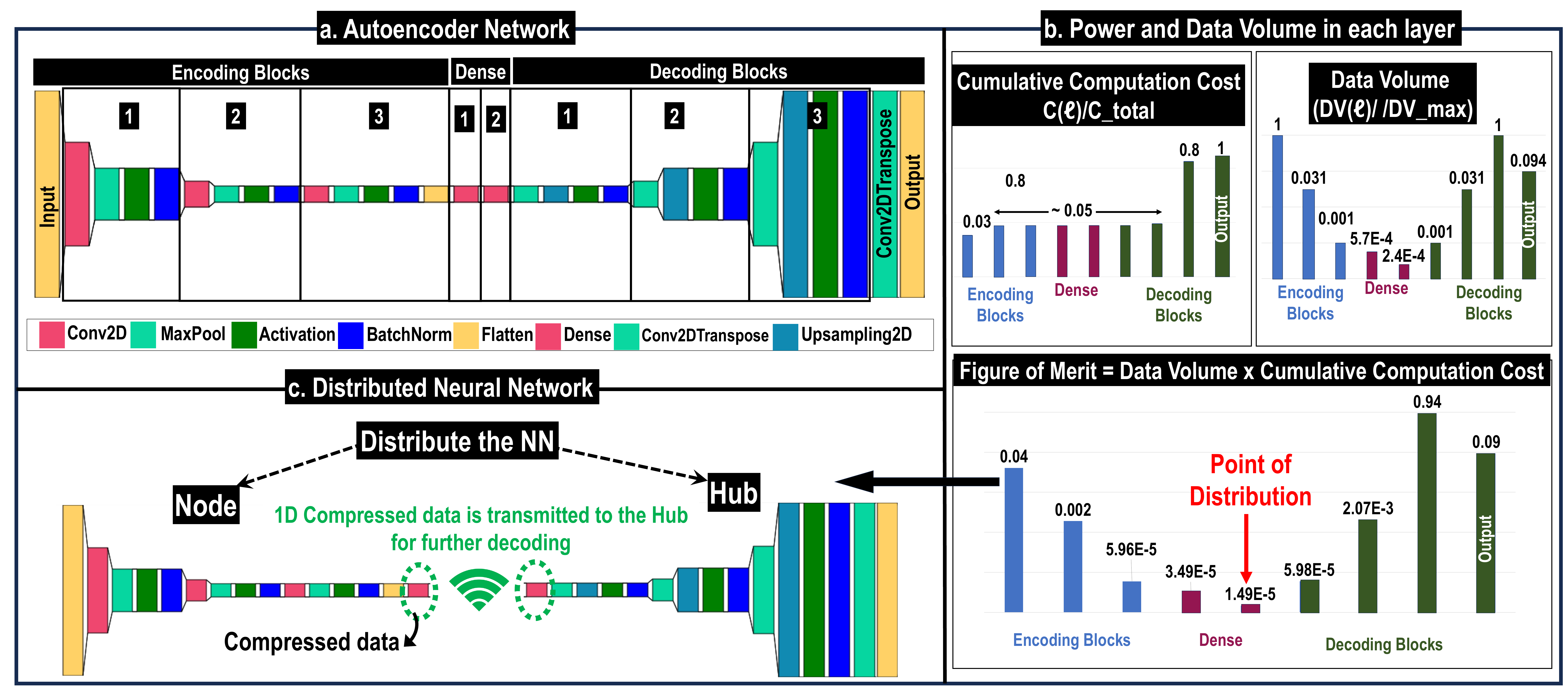}
    \caption{(a) Autoencoder architecture comprising three encoding blocks, a latent space, and three decoding blocks processing 128×128 RGB images. (b) Per-layer data volume (DV), cumulative computation cost (CC), and resulting Figure of Merit (FoM). The latent Dense layer exhibits minimum FoM, identifying the selected distribution point. FoM values are plotted un-normalized; the normalization in Eq.~\ref{eq_FoM} rescales both factors by constants and does not change the location of the minimum  (c) DistNN distribution strategy: the wearable node executes encoder layers (yellow), and the hub executes decoder layers (blue), with the distribution point in the latent space that minimizes the FoM.}
        ~\vspace{-4mm}
\label{fig_propose}
\end{figure*}

Motivated by the gaps identified in Section~\ref{sec:motivation}, we introduce DistNN, a hardware-software co-design framework for split inference, distributing neural network execution between ULP wearable nodes and resource-rich hubs. In DistNN, the node executes the early layers and transmits intermediate feature maps, while the hub completes the remaining inference or reconstruction. The main design challenge is to select a distribution point that reduces communication cost without excessive computation on the node. DistNN addresses this using a platform-independent figure of merit (FoM) based on data volume (DV) and cumulative computation cost (CC). We demonstrate the framework using AEs, whose encoder-decoder structure produces large variation in feature-map volume across layers. The same methodology also extends to other neural network architectures, as validated in Section~\ref{sec:evaluation}. This section presents the distribution strategy, energy modeling, and node-side hardware design.

\subsection{Finding the Distribution Point} \label{sec}
The distribution-point selection problem was introduced in our preliminary work~\cite{chowdhury2024leveraging}. Here, we extend that idea with an explicit problem formulation, detailed energy modeling, and a custom hardware implementation.~\vspace{-3mm}

\rev{\paragraph*{\textbf{Problem Formulation}}
For a network with $L$ candidate layers, selecting layer $\ell$ as the distribution point assigns layers $1$ through $\ell$ to the node and layers $\ell+1$ through $L$ to the hub. The transmitted data volume $DV(\ell)$ is measured at the output of layer $\ell$, i.e., at the distribution point, since only that intermediate feature map is sent to the hub. The node computation is cumulative, since the node must execute all preceding layers to produce that feature map. We define this cumulative computation cost in Eq.~\ref{eq_cc}.~\vspace{-2mm} 
\begin{equation}\label{eq_cc}
C(\ell)=\sum_{i=1}^{\ell}\#\mathrm{MAC}_i .
\end{equation}
An early distribution point may reduce node computation but require a large feature map to be transmitted, while a later point may reduce communication but increase the cumulative cost in Eq.~\ref{eq_cc}. We therefore select the layer that minimizes the combined cost $J(\ell)$, as shown in Eq.~\ref{eq_problem}. Here, the decision variable is the distribution point $\ell$, and the objective $J(\ell)$ captures the trade-off between node computation and feature-map transmission. The feasible set $\mathcal{F}$ keeps only layers that reduce the data volume relative to the input, so distribution is considered only after the network has produced a compressed representation. The following subsections define DV and CC, and then combine them into the FoM used to identify $\ell^{*}$.
\begin{equation}\label{eq_problem}
\ell^{*} = \arg\min_{\ell \in \mathcal{F}} J(\ell),
\qquad
\mathcal{F} = \{\ell \mid DV(\ell) < DV_{\mathrm{input}}\}.
\end{equation}
}\vspace{-2mm}

\begin{table}[t]
  \caption{Communication Modalities Used in the Internet of Bodies}
  \vspace{-2mm}\label{tab:comm}
  \centering
  \footnotesize
  \setlength{\tabcolsep}{5pt}
  \renewcommand{\arraystretch}{1.2}
  \begin{threeparttable}
    \begin{tabular}{@{}l l l l@{}}
      \toprule
      \textbf{Technology} & \textbf{Power} & \textbf{Throughput} & \makecell{\textbf{Energy/bit}\\\textbf{[nJ/bit]}\tnote{a}} \\
      \midrule
      BLE              & Low              & 125 kbps–2 Mbps   & 1–50 \\
      NFC              & Reader-supplied  & Up to 424 kbps    & N/A (passive)\tnote{b} \\
      Zigbee           & Very low         & Up to 250 kbps    & 5–100 \\
      Wi-Fi HaLow      & Moderate–low     & Up to $\sim$80 Mbps & 5–10 \\
      BCC/HBC         & Ultra–low        & Up to $\sim$20 Mbps & Upto 0.2\\
      LoRa             & Low              & 0.3–27 kbps       & $\sim$1{,}000–10{,}000\tnote{c} \\
      UWB              & Low–moderate     & 110 kbps–27 Mbps  & 0.01–1\\
      Backscatter (RF) & Extremely low    & Up to Gbps        & 0.0002–0.01\tnote{d} \\
      \bottomrule
    \end{tabular}
    \begin{tablenotes}\footnotesize
      \item[a] Energy/bit is strongly data-rate-, payload-, and boundary-dependent; values span TX-front-end best case to full system-level. Cross-row comparison is indicative only. \item [b] Tag energy is harvested; energy cost is borne by the reader. \item[c] System-level, spreading-factor-dependent.\item  [d] System-level, strongly tag–reader distance- and modulation-dependent
      ~\vspace{-5mm}
    \end{tablenotes}
  \end{threeparttable}
\end{table}

\subsubsection{Data Volume (DV)}
Communication energy per bit significantly exceeds computation energy per bit~\cite{maity2017wearable}, making DV minimization critical for energy-efficient distribution. Reducing transmitted DV is essential not only for energy efficiency but also for ensuring transmission reliability and data integrity. Low-power wireless protocols such as LoRa and Zigbee, while optimized for energy efficiency, are particularly susceptible to noise and electromagnetic interference. Smaller DV reduce packet loss probability and minimize energy-intensive retransmission requirements~\cite{Masoum2018Less}. Energy efficiency in wireless communication technologies varies with protocol, transmission distance, and data rate. Reported energy-per-bit values span sub-pJ/bit to µJ/bit depending on protocol, data rate, and measurement boundary, as shown in Table~\ref{tab:comm}. Efficiencies range from sub-pJ/bit for backscatter and HBC transmitter front-ends to tens of nJ/bit for standard Bluetooth, depending on operational conditions ~\cite{maity2019bodywire,Ixana81:online,OnlinePo30:online}. This range underscores the importance of managing DV to balance energy consumption with reliable communication. We calculate the DV of feature maps at layer $\ell$ using Eq.~\ref{eq_DVR}.
~\vspace{-2mm}
\begin{equation}\label{eq_DVR}
DV(\ell) = W_{\ell} \times H_{\ell} \times \#\text{Channels}_{\ell}.
\end{equation}\vspace{-4mm}

\subsubsection{Computation Cost (CC)}
The second factor is CC, which represents the arithmetic work performed on the node. In this work, CC is dominated by multiply-accumulate (MAC) operations in convolutional and dense layers. For a convolution layer, each output element is computed by applying a $k \times k$ kernel across the input channels and accumulating the products. For a dense layer, each output element is computed from the weighted sum of all input elements. For a convolution layer with kernel size $k \times k$, $C$ input channels, output dimensions $O_w \times O_h$, and $F$ filters, the MAC count is given by Eq.~\ref{eq_FLOP_a}. For a dense layer with input dimension $I_s$ and output dimension $O_s$, the MAC count is given by Eq.~\ref{eq_FLOP_b}.
~\vspace{-1mm}
\begin{subequations}
\begin{equation} \label{eq_FLOP_a}
\#\mathrm{MAC}_{\mathrm{conv}} = k^2 \cdot C \cdot O_w \cdot O_h \cdot F
\end{equation}
\vspace{-6mm}
\begin{equation} \label{eq_FLOP_b}
\#\mathrm{MAC}_{\mathrm{dense}} = I_s \cdot O_s 
\end{equation}
\end{subequations}
Pooling layers reduce the spatial dimensions of feature maps and therefore lower the MAC counts of subsequent layers, while batch normalization adds only minor arithmetic overhead during inference. The CC used in the FoM is determined primarily by convolutional and dense layers. \rev{Since the node executes every layer up to the distribution point, the relevant cost is the cumulative CC defined in Eq.~\ref{eq_cc}, rather than the MAC count of the layer $\ell$ alone.}
~\vspace{-2mm}

\rev{\subsubsection{Figure of Merit (FoM)}
Using DV and cumulative CC, we define a FoM to select the distribution point. At design time, the target hardware and communication modality may not be fixed. Therefore, the energy per transmitted bit and the energy per MAC may not be known. Instead of using hardware-specific energy values, we rank the candidate layers using only the network-level quantities DV and cumulative CC. The FoM used to rank the candidate distribution points is given in Eq.~\ref{eq_FoM}. This gives the explicit form of the objective introduced in Eq.~\ref{eq_problem}, with $J(\ell) \equiv FoM(\ell)$. The two terms are normalized, so the FoM is dimensionless and can be compared across networks of different sizes. This also addresses a limitation of our preliminary work~\cite{chowdhury2024leveraging}, where the un-normalized form was not directly comparable across architectures.
\begin{equation} \label{eq_FoM}
FoM(\ell) = \frac{DV(\ell)}{DV_{\max}} \cdot \frac{C(\ell)}{C_{\mathrm{total}}},
\qquad
\ell^{*} = \arg\min_{\ell \in \mathcal{F}} FoM(\ell).
\end{equation}
The product form makes the score depend on relative reductions in data volume and cumulative computation cost. For example, reducing either DV or cumulative CC by the same fraction has the same effect on the FoM. A weighted sum would require platform-specific energy-per-bit and energy-per-MAC values, which are not available before the hardware and communication link are chosen. The product form avoids this assumption and provides a hardware-independent ranking of candidate distribution points. The selected distribution point is the feasible layer with the lowest FoM. This point does not necessarily coincide with the layer having the minimum DV, because DV is measured only at the transmitted feature map, while CC accumulates over all node-side layers up to the distribution point. Therefore, a fixed rule such as always distributing at the latent space does not hold across architectures. The FoM is evaluated for all candidate layers using Algorithm~\ref{alg:fom}, requiring one $O(L)$ pass through the network.}

\begin{algorithm}[t]
\caption{\rev{Distribution Point Selection}}\label{alg:fom}
\begin{algorithmic}[1]
\Require $DV(\ell)$ and $\#\text{MAC}_\ell$ for $\ell=1$ to $L$; input volume $DV_{\text{input}}$
\Ensure  Distribution point $\ell^{*}$
\State $C(0)\gets 0$
\For{$\ell = 1$ to $L$}
    \State $C(\ell)\gets C(\ell{-}1)+\#\text{MAC}_\ell$ \Comment{cumulative node computation}
\EndFor
\State $DV_{\max}\gets \max_\ell DV(\ell)$, \quad $C_{\text{total}}\gets C(L)$
\State $\mathcal{F}\gets\{\ell : DV(\ell) < DV_{\text{input}}\}$ \Comment{retain only compressing points}
\State $best\gets\infty$
\For{$\ell \in \mathcal{F}$}
    \State $f\gets \dfrac{DV(\ell)}{DV_{\max}}\cdot\dfrac{C(\ell)}{C_{\text{total}}}$
    \If{$f < best$} \State $best\gets f$, \quad $\ell^{*}\gets\ell$ \EndIf
\EndFor
\State \Return $\ell^{*}$
\end{algorithmic}
\end{algorithm}

To demonstrate the methodology, we apply the FoM to the autoencoder (AE) introduced earlier. The AE processes 128$\times$128 RGB images using three encoding blocks, a latent layer, and three decoding blocks (Fig.~\ref{fig_propose}(a)). The encoder reduces the input to a compact representation, while the decoder reconstructs the image. Evaluating the FoM at each layer identifies the dense latent layer as the distribution point because it provides the best balance between feature-map volume and cumulative node computation. The node transmits this compact representation to the hub, reducing communication energy relative to raw-data offloading, and the hub executes the remaining decoder layers. Fig.~\ref{fig_propose}(b) shows the FoM across the AE layers, and Fig.~\ref{fig_propose}(c) shows the resulting distribution strategy. \revv{Within the DistNN workflow, Algorithm~\ref{alg:fom} runs once at design time to determine $\ell^{*}$. The selected point assigns the preceding layers to the ULP wearable node and the remaining layers to the resource-rich hub, while also fixing the feature-map payload transmitted between them.} \rev{All results in this work use the platform-independent FoM. When the target hardware and communication modality are known, the same candidate points can be evaluated using explicit node-energy and latency costs. The energy-aware form is $FoM_{\mathrm{E}}(\ell)=e_{\mathrm{MAC}}C(\ell)+e_{\mathrm{bit}}bDV(\ell)$, and the latency-aware form is $FoM_{\mathrm{L}}(\ell)=C(\ell)/\Theta+bDV(\ell)/R$, where $e_{\mathrm{MAC}}$ is the energy per MAC, $e_{\mathrm{bit}}$ is the energy per bit, $b$ is the activation bit width, $\Theta$ is the MAC throughput, and $R$ is the communication data rate. These forms preserve the same computation--communication trade-off while incorporating measured platform costs.} \revvv{In this work, the FoM formulation and Algorithm~\ref{alg:fom} are applied to sequential, single-path networks. The same principle can be extended to networks with skip connections, parallel branches, concatenations, or multiple heads, such as U-Net and YOLO. For these networks, a distribution point can be represented as a valid graph cut between the node and hub. The DV term would include all feature maps crossing the cut, while the computation term would count each node-side operation once. This extension requires graph-aware candidate generation and feasibility checks and is left for future work.}
~\vspace{-2mm}

~\vspace{-2mm}

\subsection{Energy Requirements for DistNN}

\label{sec:method_energy}

The FoM identifies a distribution point from DV and cumulative CC, but whether that point reduces node energy depends on the target hardware and communication link. Raw-data offloading removes node-side model computation but may incur substantial communication energy~\cite{maity2017wearable}. DistNN therefore restricts candidate distribution points to compressed representations and evaluates node compute plus communication energy against full local execution. We use the following design conditions:


\begin{enumerate}

    \item Communication-volume constraint: In this work, candidate distribution points are restricted to layers where the feature-map volume is smaller than the input, i.e., $DV(\ell) < DV_{\mathrm{input}}$. At layers where $DV(\ell) \geq DV_{\mathrm{input}}$ (for example, after transposed convolutions), transmitting the intermediate representation provides no data-volume reduction relative to raw-input offloading. The feasibility set $\mathcal{F}$ in Eq.~\eqref{eq_problem} therefore excludes such layers.
    
    \item Computational asymmetry between the node and hub: The distribution point should leave only a limited fraction of the cumulative computation on the wearable node so that its computation remains within the target ULP budget, while the remaining layers are executed on the resource-rich hub. Since this work optimizes node-side energy rather than total node-plus-hub energy, this condition is expressed in terms of the node computation budget rather than a required relation between $E_{\mathrm{node}}$ and $E_{\mathrm{hub}}$.


    \item Net node-energy saving: Distribution provides a node-side energy benefit when ($E_{\mathrm{node}} + E_{\mathrm{comm}} < E_{\mathrm{full\_local}}$). A substantially smaller value is desirable, but is not required for distribution to provide an energy benefit. Ideally, communication is also a minor share of the node budget, i.e., $E_{\mathrm{comm}} \lesssim E_{\mathrm{node}}$, although this is a preferred operating point rather than a strict requirement. Early distribution, before sufficient data-volume reduction, increases $E_{\mathrm{comm}}$ and can erode the benefit. As shown in Section~\ref{sec:evaluation}, BLE results in $E_{\mathrm{comm}} > E_{\mathrm{node}}$ for AE1, yet the total node-side energy remains below full local inference, so distribution still provides a net node-side energy saving.
\end{enumerate}
In the evaluated DistNN configurations, communication needs to be small enough that the total node energy remains below full local execution. Wi-R makes communication a minor share of the node budget, whereas BLE may have $E_{\mathrm{comm}} > E_{\mathrm{node}}$. Transmitting raw data increases communication energy in proportion to the larger payload, making raw-input offloading less favorable under the evaluated link assumptions.~\vspace{-2mm}


\begin{figure}[t]
    \centering
    \includegraphics[width=0.85\linewidth]{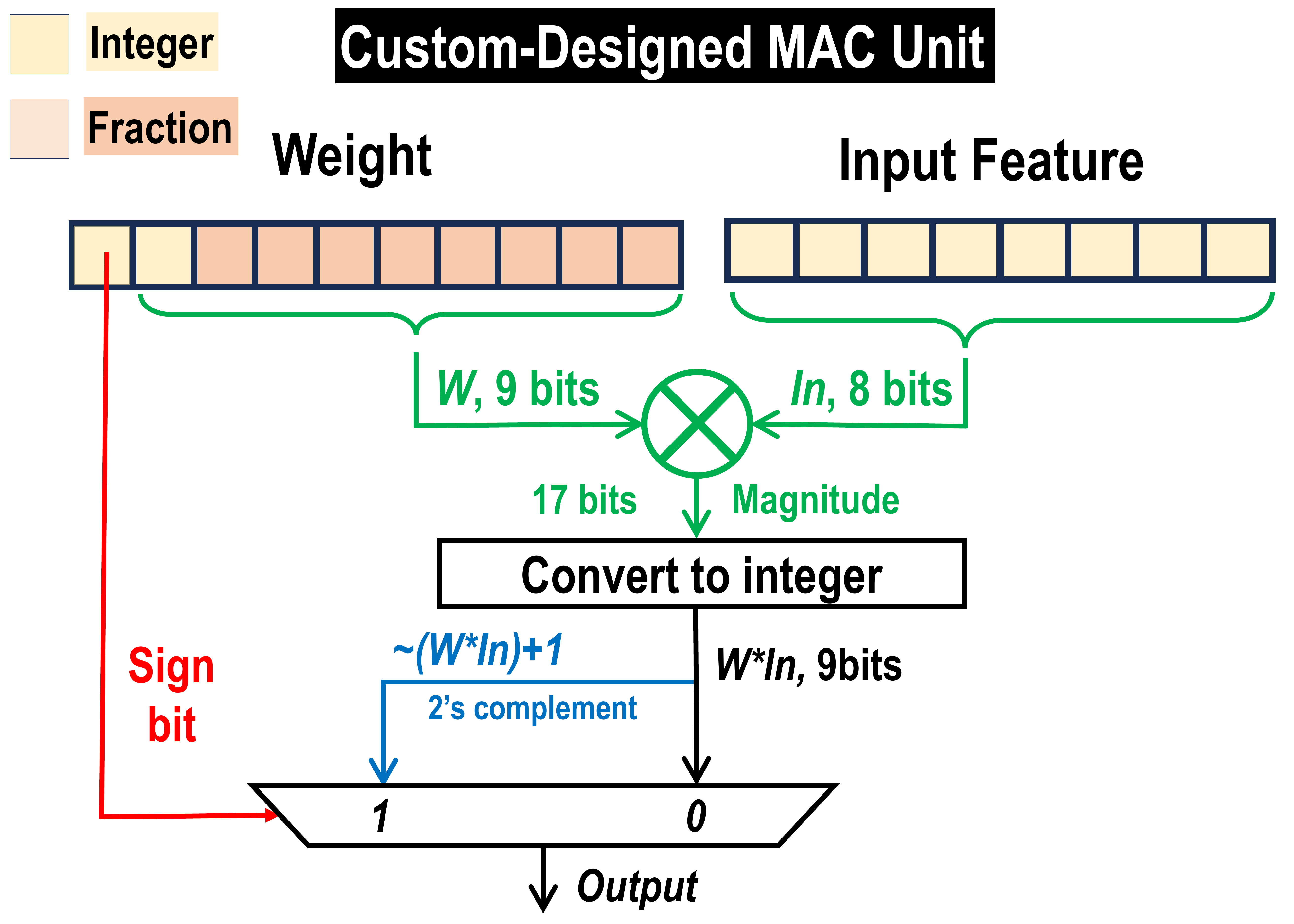}
    \caption{Custom MAC unit using fixed-point arithmetic for 8$\times$10-bit multiplication of input features (8 bits) with weights (10 bits), optimized for low-power wearable nodes.}
    ~\vspace{-4mm}
    \label{fig_mac}
\end{figure}


\begin{figure}[t]
    \centering
    \includegraphics[width=1\linewidth]{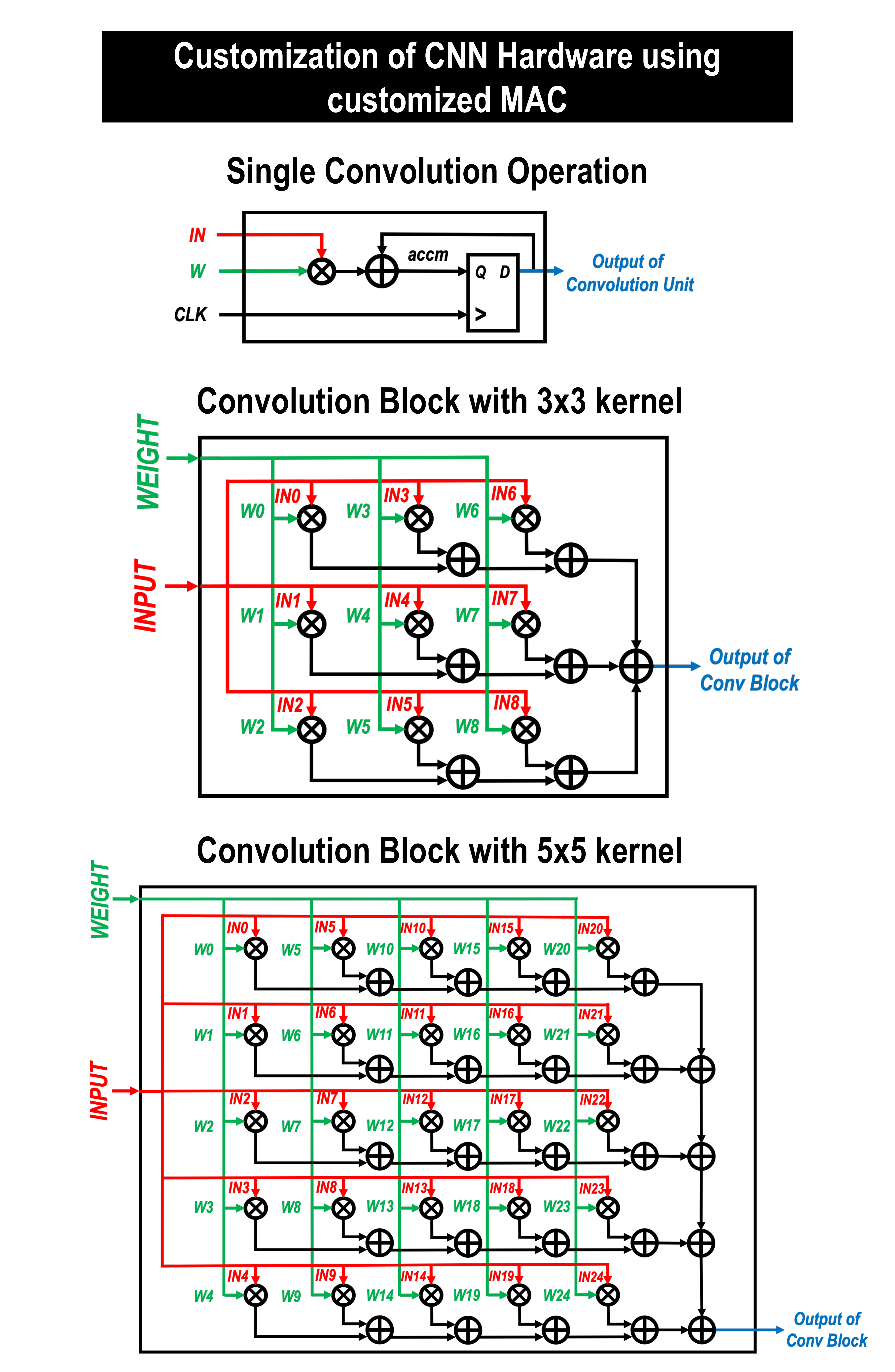} \vspace{-8mm}
    \caption{Custom-MAC-based convolution blocks. Parallel designs (conv3$\times$3\_parallel and conv5$\times$5\_parallel) use $k^2$ MAC units, while sequential designs (conv\_sequential) reuse $k$ MAC units over $k$ cycles, enabling power-latency tradeoffs for wearable nodes.}
    ~\vspace{-6mm}
    \label{fig_CNN_Verilog}
\end{figure}

\begin{table}[t]
  \centering
  \caption{Power of Different Convolution Blocks}
  \label{tab:CONV_POWER}
  \renewcommand{\arraystretch}{1.15}
  \begin{tabular}{lccc}
    \hline
    \textbf{Parallel 3x3} & \textbf{sequential 3x3} & \textbf{Parallel 5x5} & \textbf{sequential 5x5} \\ \hline
    282 uW & 93 uW & 789 uW & 157 uW  \\
    \hline
    \multicolumn{4}{l}{\footnotesize (Frequency: 100MHz, Voltage: 1V)}
    ~\vspace{-5mm}
  \end{tabular}
\end{table}
\begin{figure*}[t]
    \centering
    \includegraphics[width=0.98\linewidth]{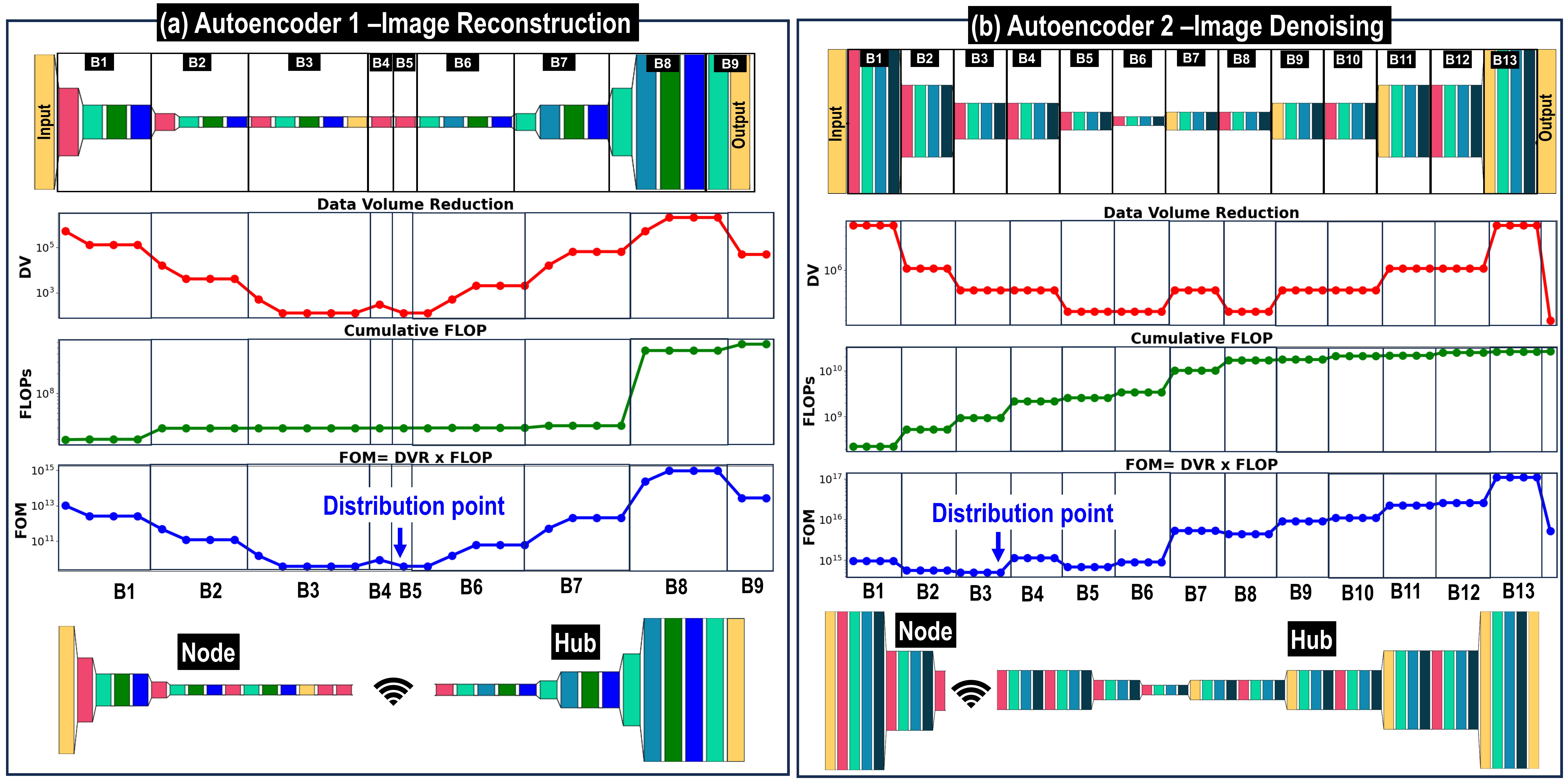}
    \caption{FoM-based distribution point identification for autoencoder architectures. (a) AE1 for image reconstruction shows a distribution at B5. (b) AE2 for image denoising identifies distribution at B3 (an earlier encoding layer), demonstrating that FoM adapts to architectural variations and may select distribution points outside the latent space depending on layer-wise computation-communication tradeoffs.}
    ~\vspace{-4mm}
    \label{fig_res}
\end{figure*}

\subsection{Application-Specific ML Hardware for ULP Operation}
\label{ulp}

The distribution strategy and energy modeling framework presented so far establish the requirements for wearable node operation. Achieving ULP ($\leq$10mW) targets derived in Section~\ref{sec:method_energy} requires hardware specifically optimized for wearable constraints rather than general-purpose processors. As discussed in Section~\ref{sec:lit}, traditional ML hardware such as GPUs or generalized ML accelerators are power and memory intensive. These platforms typically use 32-bit IEEE 754 floating-point arithmetic for flexibility and accuracy, but such precision is not compatible with the tight energy and memory limits of wearable devices. For example, the NVIDIA GeForce RTX 3060 and RTX A5000 have maximum power draws of 170 W and 230 W, respectively~\cite{GeForceR8:online,Professi52:online}, which is impractical for battery-powered wearable systems.

To meet these constraints, we adopt an application-specific ML hardware design tailored to wearable nodes. The key idea is to use low-bitwidth fixed-point arithmetic in place of 32-bit floating point, thereby reducing both memory footprint and computational cost while maintaining adequate precision for the target workloads. At the core of this design is a custom multiply–accumulate (MAC) unit, which serves as the fundamental building block for convolutional and dense layers. As illustrated in Fig.~\ref{fig_mac}, weights are represented using a 10-bit fixed-point format (1 sign bit, 1 integer bit, and 8 fractional bits), supporting values in the range \([-1,1]\). Feature maps are represented using 8-bit integers, consistent with typical RGB input ranges.The MAC computes the product, applies sign correction via two’s complement when required, and accumulates into a wider register as the network deepens, causing feature-map bitwidths to grow up to 16 bits.
\revvv{Specifically, the sign bit is stored separately from the 9-bit weight magnitude, so the multiplier itself operates on unsigned operands (9-bit magnitude x 8-bit feature), producing a 17-bit product before it is reduced to a 9-bit result. A 2-to-1 multiplexer, controlled by the sign bit, then selects between this result and its two's-complement negation, $\sim$(W x In)+1, to apply sign correction only once, at the output.}


The MAC design was implemented in Verilog and synthesized using Synopsys Design Compiler in a 65~nm TSMC CMOS process. At a 100~MHz clock frequency and 1~V supply, the custom MAC achieves an energy of 230~fJ per operation. Using Eq.~\ref{eq_pwr} together with the MAC counts from Eqs.~\eqref{eq_FLOP_a} and~\eqref{eq_FLOP_b}, we can directly estimate each layer's contribution to node-side power and verify that the overall computation fits within the ULP budget.
~\vspace{-2mm}
\begin{equation}\label{eq_pwr}  
\mathrm{Power} = \#\mathrm{MAC\ /frame} \times (\mathrm{Energy/MAC}) \times \mathrm{fps},
\end{equation} 
~\vspace{-4mm}

Beyond the scalar MAC, we also validate its use in standard convolutional blocks through synthesis in Synopsys Design Compiler. Figure~\ref{fig_CNN_Verilog} shows the schematic of 2D convolution using the custom MAC as a primitive. A basic convolution unit consists of one MAC and an accumulator, which is spatially replicated to construct convolution blocks for kernels of size \(k \times k\), where \(k \in \{3,5\}\) in our experiments. For each kernel size, we realize two microarchitectures: a \emph{parallel} implementation, in which all \(k^2\) MAC units are executed in a single cycle to compute one output window per clock, and a \emph{sequential} implementation, in which \(k\) MAC units are reused over \(k\) cycles to compute the same window. 
Both designs perform the same \(k^2\) MAC operations per window (25 MACs for 5x5, 9 MACs for 3x3, a 2.78x ratio), so the total energy per window, which is power times cycle count, scales directly with MAC count. The parallel design completes a window in one cycle, so its energy ratio equals its power ratio; the sequential design runs k cycles per window, so its energy ratio is recovered only after multiplying the power by the cycle count. Under 100 MHz and 1 V, the Conv2D-3×3 block consumes 282 µW in the parallel implementation and 93 µW in the sequential implementation.  For the Conv2D-5×5 block, the parallel and sequential implementations consume 789 µW and 157 µW, respectively. The parallel power ratio (789/282 $\approx$ 2.80) already matches the 2.78x MAC ratio because each window takes one cycle. The sequential power ratio (157/93 $\approx$ 1.69) is smaller, but scaling by the k = 5 and k = 3 cycle counts recovers an energy ratio of (157 × 5)/(93 × 3) $\approx$ 2.81, again matching the MAC ratio. Synthesized RTL results, including routing, control, and an 80\% memory-activity overhead~\cite{verma2019memory}, show that both implementations remain within the ULP ($\leq$10mW) regime. These results confirm that standard CNN building blocks can be efficiently realized using the proposed MAC and that different power–latency points can be supported on the wearable node. The datapath processes convolutions in a streaming, line-buffer fashion, wherein only the k input rows required by the active kernel window are buffered (approximately k x W x C values, under 50~KB for AE2's first encoder layer at 8-bit precision), and output rows are streamed onward as they complete. Full activation tensors are therefore never materialized on the node, in contrast to interpreter-based MCU deployment, where the TFLite Micro arena must hold the complete peak tensor (Section ~\ref{sec:rq3_system}).



%% file: 4_eval_rev.tex
\section{Evaluation}
\label{sec:evaluation}

Having outlined the architecture and design principles of DistNN, we now evaluate it through experiments addressing key feasibility and performance questions. This evaluation focuses on the wearable node, the critical energy bottleneck in distributed systems, since hub-side processing leverages existing high-performance computing infrastructure (smartphones, edge servers) with abundant energy resources and established GPU optimization techniques. \revvv{Published measurements place this hub class in the watt regime, two to four orders of magnitude above the node budget targeted.} Our contributions center on demonstrating that ULP node operation is achievable through hardware-software co-design, enabling always-on wearable intelligence previously infeasible with prior on-device execution and not addressed by prior distributed approaches within wearable memory and ULP budgets.

We address the following research questions covering the FoM, feasibility, accuracy, and energy:

\begin{enumerate}
\item \textbf{Scalability of FoM-Based Distribution:} How effectively does the FoM-based distribution strategy identify suitable distribution points across different AE architectures, using only network topology?


\item \textbf{Distribution Necessity and Node Energy:} Can wearable-class microcontrollers sustain full-model inference locally, or is distributed execution fundamentally required to meet resource constraints?

\item \textbf{Impact of Numerical Precision on DistNN Performance:} How does the custom hardware implementation improve energy efficiency while preserving output quality for ULP wearable nodes?

\item \textbf{System Level Validation of DistNN:} How does DistNN's node implementation compare to state-of-the-art ML accelerators and distributed inference approaches in terms of energy efficiency for wearable-scale operation?
\end{enumerate}


\subsection{Scalability of FoM-Based Distribution Strategy} 
\label{eval_scalability}

We evaluate the scalability of our FoM-based distribution methodology using two typical machine-vision workloads, namely, image reconstruction and image denoising. Both rely on autoencoders, which are challenging to distribute because their data volume decreases through the encoder and then expands through the decoder. This non-monotonic pattern creates several viable distribution points across the architecture, unlike classification CNNs, where spatial dimensions only shrink. Because of this variability, selecting a distribution point cannot rely on heuristics. Our FoM explicitly computes the tradeoff between data volume and computational load at each candidate layer, and the resulting value identifies the distribution point that minimizes the FoM for that architecture. We examine two distinct use cases (image reconstruction and image denoising) demonstrating how FoM identifies distribution points for networks with different depths and reconstruction objectives.

\subsubsection{\textbf{AE1: Image Reconstruction}}
This is an image reconstruction AE using the CelebA dataset~\cite{CelebADa70:online}. As illustrated in Fig.~\ref{fig_res}(a), data volume undergoes reduction through the encoder until reaching the feature maps representation in Block B5, then increases through the decoder until the final block B9. The number of FLOPs surges at the decoding block B8 due to the expansion of the output feature map dimensions. Based on the FoM defined in Eq.~\eqref{eq_FoM}, we determine the distribution point at the Dense Layer B5, where the FoM is minimized: it combines low data volume with modest cumulative computation, enabling efficient feature-map transmission.

\subsubsection{\textbf{AE2: Image Denoising}}
Fig.~\ref{fig_res}(b) shows an image-denoising autoencoder with 13 blocks (5 encoding layers, one dense layer, and seven decoding layers), trained using noisy images from~\cite{UCBerkel57:online}. The model processes 256×256×3 inputs to produce denoised outputs of identical dimensions. Based on FoM, the  distribution point is located at block B3, an earlier encoding layer rather than the latent space. This shows that the FoM-selected distribution point depends on architecture-specific tradeoffs and need not reside in the latent space.

These results demonstrate that the FoM identifies effective distribution points across AEs of different depth (9 vs. 13 blocks), latent representation (dense vs. convolutional), and application (reconstruction vs. denoising). Users can adjust the distribution point for energy or latency specific constraints, but the FoM provides a balanced default by minimizing the product of data volume and cumulative computation cost, trading off node computation against communication overhead. Sections~\ref{res_pwr} and~\ref{sec:rq3_system} validate that the FoM-based distribution enable ULP node operation while maintaining distributed execution feasibility.  \textit{Note:} While this evaluation focuses on autoencoder architectures, Section~\ref{res_pwr} validates that the custom hardware implementation extends to classification CNNs (VGG, EfficientNet on CIFAR-100). 
\vspace{-4mm}

\begin{figure}[t]
  \centering
  \includegraphics[width=0.8\columnwidth]{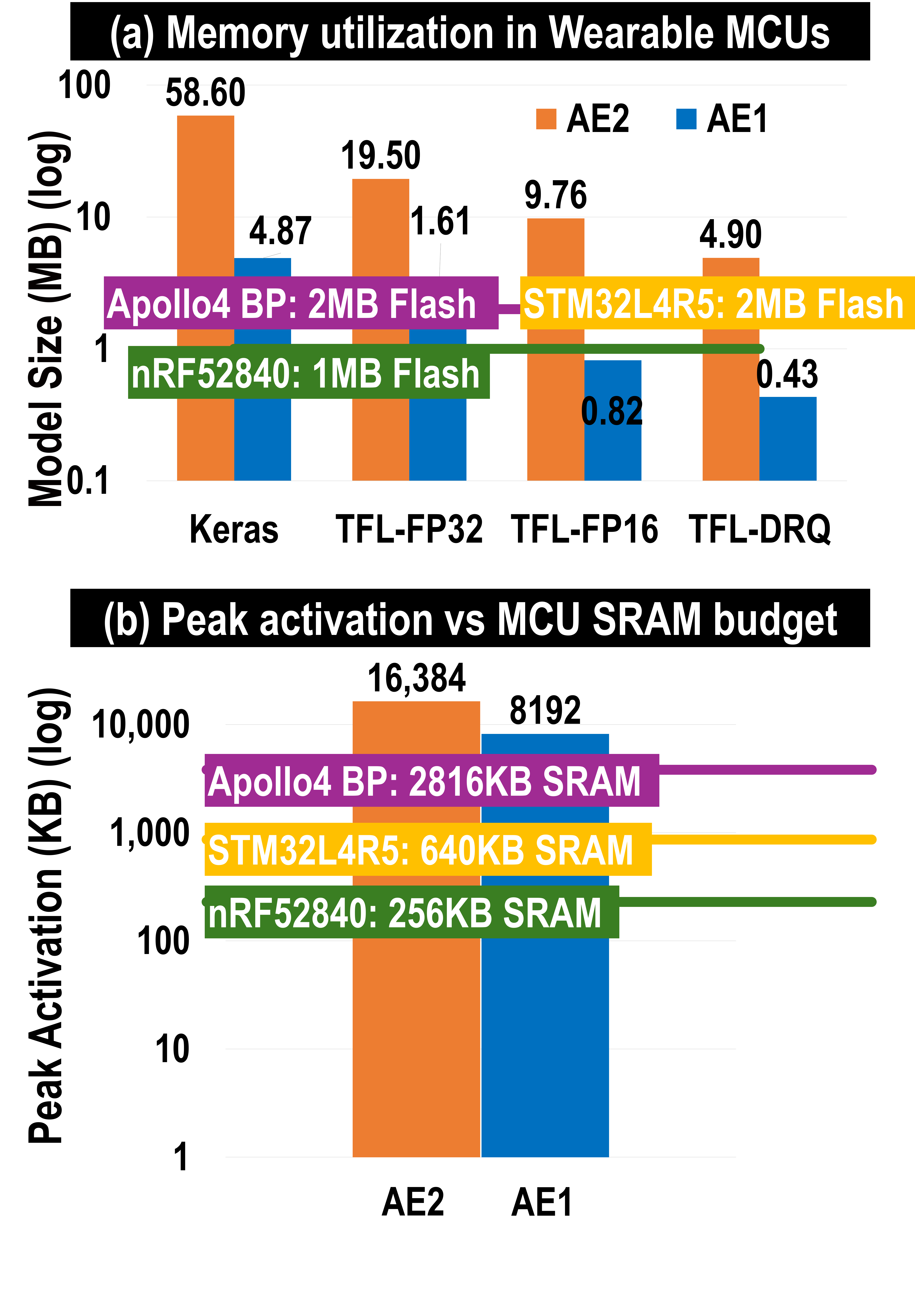}
  \vspace{-1cm}
\caption{\rev{Memory utilization in wearable MCUs for AE1 (orange) and AE2 (blue) under different TFLite configurations. (a)~Model size vs. Flash capacities of nRF52840, STM32L4R5, and Apollo4 Blue Plus. (b)~Peak activation vs. SRAM capacities of the same devices}.
~\vspace{-4mm}}
  \label{fig:iot_memory}
\end{figure}

\rev{
\subsection{Necessity of DistNN with Wearable MCU Validation}
\label{sec:rq3_system}
Having established that the FoM identifies the distribution point across architectures, we now ask whether such distribution is fundamentally required, or whether the complete neural network models could instead execute directly on a typical wearable microcontroller using TinyML deployment strategies. We consider AE1 and AE2 from Section~\ref{eval_scalability} and compare their memory requirements with those of three wearable-class microcontrollers spanning the realistic deployment envelope for on-body sensing nodes: \textbf{(i) }the Nordic nRF52840 (Cortex-M4F, 1~MB Flash, 256~KB SRAM)~\cite{nordic_nrf52840} which powers the Arduino Nano~33~BLE Sense, \textbf{(ii) } the STMicroelectronics STM32L4R5 (Cortex-M4, 2~MB Flash, 640~KB SRAM)~\cite{st_stm32l4r5} \textbf{(iii) }the Ambiq Apollo4 Blue Plus (Cortex-M4F, 2~MB MRAM, 2.75~MB SRAM)~\cite{ambiq_apollo4bp}.  The nRF52840 and STM32L4R5 are designated reference platforms of the MLPerf Tiny benchmark suite~\cite{banbury2021mlperf}, while the Apollo4 Blue Plus is representative of the highest-end wearable application processors currently in production. In the context of DistNN, these devices represent the class of \emph{resource-constrained nodes} on which on-body computation would be deployed. 
The original Keras implementations of AE2 and AE1 store 32-bit floating-point (FP32) weights and include training-related metadata, optimizer state, and graph-construction objects. TensorFlow Lite (TFLite), in contrast, packages the inference graph into a compact FlatBuffer representation that removes this auxiliary state, reducing the binary size even without changing numeric precision. We evaluate three TFLite configurations commonly used in embedded TinyML deployments:
\begin{itemize}
    \item TFL-FP32 (no optimization): FP32 weights retained; size reduction arises solely from the FlatBuffer representation.
    \item TFL-FP16 (weight quantization): weights stored in 16-bit floating point with activations remaining floating point.
    \item TFL-DRQ (dynamic-range quantization): weights quantized to INT8, while activations remain floating point to avoid full
    integer calibration.
\end{itemize}
Monolithic deployment requires satisfying two memory constraints simultaneously. The serialized model must fit in non-volatile memory storage (Flash or MRAM), and the largest activation tensor produced during inference must fit in SRAM. We evaluate both. For AE2, the 4 configurations yield model sizes of 58.595~MB (Keras), 19.503~MB (TFL-FP32), 9.758~MB (TFL-FP16), and 4.904~MB (TFL-DRQ). For AE1, the corresponding sizes are 4.873~MB, 1.609~MB, 0.819~MB, and 0.433~MB. Figure~\ref{fig:iot_memory}(a) plots these against the Flash capacities of the three reference devices. Every AE2 variant exceeds every device's Flash budget. Even at TFL-DRQ, AE2 is roughly $5\times$ the nRF52840 budget and $2.5\times$ that of the STM32L4R5 and Apollo4~BP. AE1 fares better at the Flash level, where the TFL-DRQ AE1 ($0.433$~MB) fits within the Flash of all three devices. However, flash capacity alone is therefore not a sufficient feasibility criterion. The binding constraint is SRAM. Hence, we also profile the forward pass of each model and report the size of the largest single activation tensor as a lower bound on the TFLite Micro arena that must be allocated at inference time~\cite{david2021tflm}. For AE2, the peak occurs at the first encoder convolution (256$\times$256$\times$64), requiring 16{,}384~KB at FP32. For AE1, the peak occurs in the decoder upsampling stage (128$\times$128$\times$128), requiring 8{,}192~KB at FP32. Fig.~\ref{fig:iot_memory}(b) shows these against the SRAM capacities of the same three devices. Both peaks exceed every SRAM budget at every TFLite precision considered. AE2's peak activation is $64\times$ the nRF52840 SRAM budget, $25.6\times$ that of the STM32L4R5, and $5.8\times$ that of the Apollo4~BP. AE1's peak is $32\times$, $12.8\times$, and $2.9\times$ the same budgets, respectively. The peak activation memory is invariant under TFL-FP16 and TFL-DRQ because both configurations quantize weights only; activation tensors remain in floating point.
More aggressive compression techniques like full INT8 activation quantization, pruning, or weight clustering, could, in principle, further reduce activation footprint by approximately $4\times$. However, such techniques degrade reconstruction fidelity, as shown in Section~\ref{res_pwr}. Even with full INT8 activations, AE2 would still exceed the Apollo4~BP SRAM budget by $1.45\times$, and the STM32L4R5 budget by $6.4\times$, and AE1 would still exceed the STM32L4R5 budget by $3.2\times$. Meeting the memory budgets through additional compression, therefore, violates quality requirements without resolving the infeasibility on the MLPerf Tiny reference platforms. 
A possible workaround to the SRAM bound is to store the activation tensors in an on-device non-volatile memory and reload them when needed. However, this introduces a tradeoff between SRAM space and extra memory access. For example, the nRF52840 writes flash one 32-bit word at a time, spending 41~\textmu s per write with the CPU halted~\cite{nordic_nrf52840}. Even using the device's CPU-active current as a conservative proxy (3.1--6.3~mA at 3.0~V~\cite{nordic_nrf52840}), this implies roughly 12--24~nJ/bit, consistent with measured embedded-flash write energies on the order of 16~nJ/bit~\cite{mathur2006}. A BLE transmission costs approximately 19~nJ/bit and an HBC link approximately 0.12~nJ/bit~\cite{Ixana81:online}. On-device flash storage is therefore comparable in energy to BLE transmission and roughly two orders of magnitude more costly than the HBC radio~\cite{mathur2006}, eliminating any energy advantage from local storage. Write throughput is a further limitation. The 41~\textmu s per-word write timing caps the effective rate at approximately 0.78~Mbit/s, so a multi-megabyte activation tensor requires several seconds to store. The STM32L4R5 is similarly constrained, with a sustained write rate below 100~KB/s~\cite{st_stm32l4r5}. Flash endurance compounds both issues. NOR flash sustains on the order of $10^{4}$ write cycles~\cite{nordic_nrf52840, st_stm32l4r5}, so spilling activations once per frame would exhaust the device within hours of continuous operation. The Apollo4 Blue Plus replaces flash with MRAM, raising endurance beyond $10^{14}$ cycles~\cite{lee2022mram}. However, at  $\sim$25~pJ/bit per write and a memory architecture provisioned for code storage rather than sustained streaming writes~\cite{ambiq_apollo4bp}, per-frame activation storage remains infeasible. Since neither AE2 nor AE1 can satisfy both Flash and SRAM constraints on realistic wearable-class nodes at any precision compatible with reconstruction quality, monolithic TinyML execution of the full autoencoder is not a viable design option. \textit{\textbf{Distribution is therefore not an optimization preference but a deployment requirement, and DistNN enables these models to operate on embedded sensing nodes by keeping only a compact encoder on the device, thereby reducing on-node storage by more than an order of magnitude while enabling reconstruction quality to be recovered by the hub decoder.}}
~\vspace{-2mm}
}



\begin{table}[t]
  \centering
  \caption{Impact of Numerical Precision on Performance on NN}
  \label{fig_power}
  \renewcommand\arraystretch{1.5}
  \setlength{\tabcolsep}{5pt}
  \begin{tabular}{p{1.8cm} p{1.6cm} p{1cm} p{1cm} p{1.1cm}}
    \hline

    \textbf{Task} & \textbf{Network} & \textbf{W, FM: FP32} & \textbf{W:F10; FM:FP32} & \textbf{W:F10; FM:F8} \\ \hline

    \multirow{3}{=}{CIFAR-100 Classification}
      &            & \multicolumn{3}{c}{\textbf{Error \%}} \\ \cline{3-5}
      & VGG        & 12.15\% & 12.24\% & 13.7\%  \\ \cline{2-5}
      & EfficientNet & 9.28\%  & 9.97\%  & 13.45\% \\ \hline

    \multirow{3}{=}{Image Manipulation (AE)}
      &            & \multicolumn{3}{c}{\textbf{SSIM}} \\ \cline{3-5}
      & Reconstruction & 0.836 & 0.827 & 0.675 \\ \cline{2-5}
      & Denoising      & 0.898 & 0.882 & 0.678 \\ \hline
  \end{tabular}
  \vspace{2pt}
  \parbox{\columnwidth}{\raggedright\footnotesize
  W: Weight; FM: Feature Map; FP32: Floating Point 32 bits; F10: Fixed Point 10 bits;
  F8: Fixed Point 8 bits. Classification results show error (lower is better); image
  manipulation results show SSIM (higher is better).}
  \vspace{-4mm}
\end{table}

\subsection{Impact of Numerical Precision on Overall Performance}
\label{res_pwr}

In this section, we validate our design choice of low-precision custom hardware (10-bit weights, 8-bit feature maps) introduced in Section~\ref{ulp} to determine whether it maintains acceptable output quality across different task types. We evaluate the impact of reduced numerical precision on both classification and image manipulation performance, and compare this fixed-point representation with standard 32-bit floating-point. The first task is CIFAR-100 classification using VGG and EfficientNet CNNs; the second is autoencoder-based image reconstruction and denoising. This evaluation demonstrates that the low-precision node hardware approach extends across different computer-vision task types.

Table~\ref{fig_power} quantifies the accuracy and quality tradeoffs across precision levels. For classification networks, reducing the precision marginally affects accuracy only when the bit-depth is insufficient. VGG exhibits classification error increasing from 12.15\% (FP32) to 12.24\% (10-bit weights with FP32 features) to 13.7\% (10-bit weights with 8-bit features). EfficientNet shows similar trends with errors of 9.28\%, 9.97\%, and 13.45\%, respectively. These results demonstrate that 10-bit fixed-point weight representation provides acceptable accuracy for classification tasks, with more pronounced degradation only when both weights and feature maps use reduced precision. For autoencoder reconstruction tasks, we use Structural Similarity (SSIM) as the evaluation metric. SSIM measures perceptual similarity by comparing luminance, contrast, and structural information across local image patches, and reports quality on a 0–1 scale. Transitioning from FP32 to 10-bit weight precision (while keeping feature maps in FP32) leads to minimal degradation (0.836→0.827 for reconstruction and 0.898→0.882 for denoising). Reducing feature-map precision to 8-bit introduces larger standalone drops, to 0.675 and 0.678, indicating that reconstruction quality is more sensitive to feature-map precision than to weight precision. As we show next, this node-side degradation is recovered by the hub-side decoder, which is the basis of DistNN's asymmetric design.

\begin{figure}[t]
    \centering
    \includegraphics[width=1\linewidth]{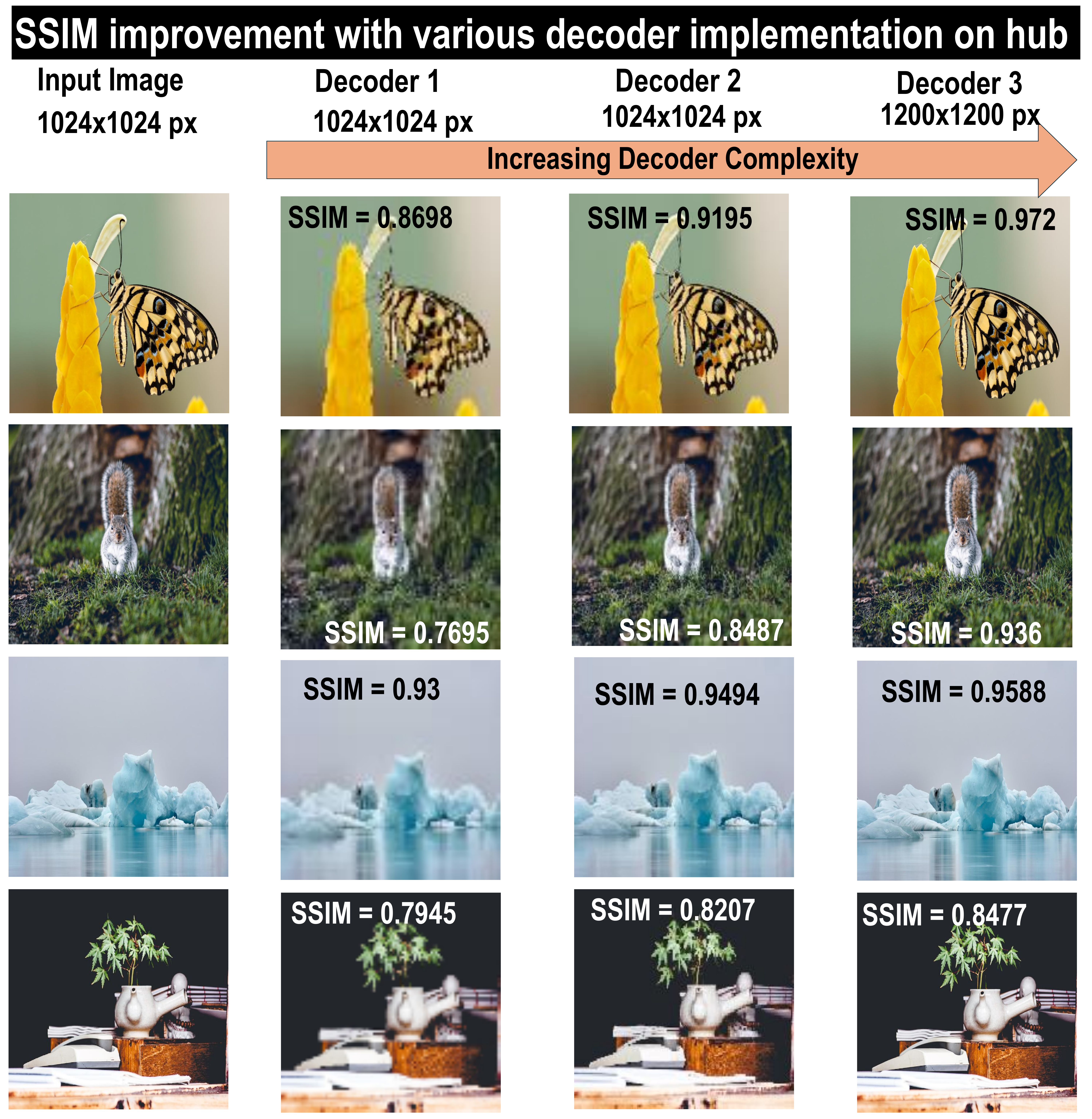} \vspace{-8mm}
    \caption{Final reconstruction quality versus hub-side decoder complexity, using identical 8-bit node-side features. As decoder complexity increases (Decoder 1 $\rightarrow$ Decoder 3), SSIM improves across all test images, reaching up to 0.97 with the most complex decoder. This demonstrates that quality lost to low-precision node computation is recovered at the resource-rich hub, supporting DistNN's asymmetric node-hub design.}
    ~\vspace{-5mm}
    \label{fig_SR}
\end{figure}

As illustrated in Fig.~\ref{fig_SR}, DistNN can run the node at low precision for ULP operation and still recover high quality at the hub. The node-side features are held fixed at the 8-bit representation that enables ULP operation, and the hub then applies decoders of increasing complexity to those same features. As decoder capacity grows, SSIM on the shown examples rises from an average of approximately 0.84 to approximately 0.93, reaching up to 0.97, all without changing anything on the node. This is the core of DistNN’s asymmetric design. The node uses low-precision feature extraction to reduce node-side energy, while the resource-rich hub performs higher-complexity decoding. Low node precision therefore provides a node-side power advantage, while additional hub-side decoding can recover part of the associated quality loss without additional node-side computation.

\textit{\textbf{Overall, these results show that DistNN maintains classification accuracy close to 32-bit floating point at the node, while reaching an average SSIM of approximately 0.90 (up to 0.97) for reconstruction. The node retains only the energy-efficient fixed-point units and ULP reported in Section~\ref{sec:method_energy}, so the quality recovery does not require additional node-side computation. By combining a low-precision fixed-point node with a higher-capacity hub decoder, DistNN preserves E2E output quality under tight node energy constraints.
}}
~\vspace{-2mm}

\rev{\subsection{System-Level Validation of DistNN}} \label{sec_eval_sys_val} \vspace{-1mm}

Having shown that the platform-independent FoM locates distribution points across architectures, that the custom fixed-point hardware helps to enable $<10$mW node operation, and that memory constraints in typical wearable MCU nodes make distribution necessary, we now position DistNN in the ML hardware landscape and validate the full pipeline end-to-end. We first compare node-side energy against accelerators spanning datacenter GPUs to ULP edge platforms, which shows where the wearable-inference regime sits and what sets it apart; node-side energy here includes both compute and the radio, for which we evaluate two most commonly used low-power modalities, the BLE standard and Wi-R, an emerging human body communication (HBC) link (Table~\ref{tab:comm}). We then estimate end-to-end latency per inference.

\paragraph{\textbf{Node-side Energy}}
For node compute, we use reported per-MAC energy for NVIDIA GPUs and fused-MAC designs~\cite{arunkumar2017mcm,park20219}, and power estimates from reported specifications for recent ML ASICs~\cite{ottavi2023dustin,miro2022samurai,ju2022systolic,garofalo2022darkside,rossi2021vega,nadalini20233}. DistNN uses our measured $0.23$~pJ/MAC efficiency with a $1.8\times$ factor for memory-access energy. Table~\ref{tab:summary} reports both energy per inference ($E$, mJ) and power ($P$, mW). Energy per inference is the rate-independent cost of one forward pass and is our primary metric \& power is the corresponding average draw at an assumed continuous 30~fps. \rev{GPU-class devices operate at $\sim\!200$~mW and above and are built for throughput and workload flexibility in datacenter and edge-server contexts. The $\sim$3 orders of magnitude gap reflects the very different design regimes of the two platforms and should not be interpreted as a like-for-like efficiency comparison.} The closest comparison set is ML ASICs ($1$--$54$~mW) and ULP edge platforms such as Vega ($1.25$--$1.3$~mW) and Tiny-Vers ($7$~mW), which reach the embedded range through domain-specific specialization but do not address this workload class within wearable memory budgets. \rev{The closest prior work is Category~4 distributed inference (Section~\ref{sec:motivation}) which includes DISSEC~\cite{li2022dissec} that runs distributed execution on Raspberry~Pi clusters, and DDSNN~\cite{bolat2025decentralized} partitions LeNet across ESP32-S3 microcontrollers (512~KB SRAM) at 
170--377~mW. Both rely on algorithmic partitioning of general-purpose devices for classification, so neither reaches the regime DistNN occupies. DistNN's compute-only power is $0.204\,\mathrm{mW}$ for AE1 and $6.54\,\mathrm{mW}$ for AE2. For comparison with prior node values that include communication, the corresponding DistNN totals with Wi-R are $0.208\,\mathrm{mW}$ and $8.43\,\mathrm{mW}$, respectively. This comes from three separate choices. First, the custom 65\,nm fixed-point MAC at $0.23$~pJ/MAC runs one to two orders of magnitude below the per-operation energy of a general-purpose MCU instruction stream, which accounts for most of the gap to MCU-based systems such as DDSNN. Second, 10-bit weight \& 8-bit feature-map fixed-point arithmetic gives the standard reduction over FP32 and a further reduction over int8 software execution. Third, the FoM distribution-point selection sets where in the network these savings are spent. A comparison on identical silicon is not feasible, because the AE workloads exceed the on-node memory of the target microcontrollers (Section~\ref{sec:rq3_system}) and re-fabricating prior designs in 65\,nm is out of scope. Also, the memory budget of DistNN is at least an order of magnitude tighter, ruling out any approach that holds the full model on-node; and AE has expansion and compression, inverting the communication-cost trade-off that classification-oriented partitioning assumes. Addressing both is what lets DistNN operate in the ULP ($\leq$10\,mW) regime on workloads that prior accelerators in the same power envelope do not target.}

\begin{table}[t]
  \centering
  \caption{\rev{ML hardware design space for wearable node implementation.}}
  \vspace{-2mm}
  \label{tab:summary}
  \renewcommand\arraystretch{1.3}
  \setlength{\tabcolsep}{3.5pt}
  \scriptsize
  \resizebox{\columnwidth}{!}{%
  \begin{tabular}{l c c c c c}
    \hline
     & \multicolumn{2}{c}{\textbf{AE1}} & \multicolumn{2}{c}{\textbf{AE2}} & \textbf{Impr.} \\
    \cline{2-3}\cline{4-5}
    \textbf{Hardware} & \textbf{P (mW)} & \textbf{E (mJ)} & \textbf{P (mW)} & \textbf{E (mJ)} & \textbf{($\times$)} \\ \hline
    \textbf{OURS}                              & \textbf{0.204} & \textbf{0.007} & \textbf{6.54}     & \textbf{0.218} & \textbf{1.00} \\ \hline
    NVIDIA MCM GPU~\cite{arunkumar2017mcm}     & 222.46 & 7.415 & 7{,}125.84 & 237.53 & 1089.41 \\
    Tiny Vers~\cite{jain2023tinyvers}          & 6.99   & 0.233 & 223.90     & 7.46   & 34.23 \\
    Darkside~\cite{garofalo2022darkside}       & 4.29   & 0.143 & 137.47     & 4.58   & 21.02 \\
    Dustin~\cite{ottavi2023dustin}             & 1.79   & 0.060 & 57.27      & 1.91   & 8.76 \\
    ASIC (Fused Opt.\ MAC)                     & 1.69   & 0.056 & 54.03      & 1.80   & 8.26 \\
    Samur AI~\cite{miro2022samurai}            & 1.32   & 0.044 & 42.41      & 1.41   & 6.48 \\
    Vega~\cite{rossi2021vega}                  & 1.25   & 0.042 & 40.15      & 1.34   & 6.14 \\
    SNCPU~\cite{ju2022systolic}                & 0.644  & 0.021 & 20.62      & 0.687  & 3.15 \\
    Flex V~\cite{nadalini20233}                & 0.209  & 0.007 & 6.68       & 0.223  & 1.02 \\ \hline
  \end{tabular}}
  \parbox{\columnwidth}{\raggedright\footnotesize
Per-node power and energy processing AE1/AE2 at 30~fps. OURS is node-compute only. Impr.\ is the power/energy ratio to OURS.} \vspace{-5mm}
\end{table}

\rev{For end-to-end node power, we need to add the communication cost to the total node power. Taking AE1 as an example, the node transmits a $128$-dimensional feature vector at B5 (8-bit), i.e.\ $1024$~bits per frame, which at 30~fps is a $30{,}720$~bits/s stream. Using the in-production YR23 Wi-R chip at $0.12$~nJ/bit~\cite{Ixana81:online}, draws $30{,}720 \times 0.12~\text{nJ/bit} \approx 3.7~\mu$W, for a total of $\approx\!0.208$~mW. BLE, at a representative transmitter energy of $19$~nJ/bit from the nRF52840 datasheet (TX current at $0$~dBm and $1$~Mbps)~\cite{nordic_nrf52840}, draws $\approx\!0.584$~mW, for a total of $\approx\!0.788$~mW. Wi-R meets the ideal case $E_{\text{comm}}\ll E_{\text{node}}$ by more than $50\times$, while BLE sits at $\approx\!2.9\times$ node compute power. For AE2, the Wi-R radio draws $\approx\!1.89$~mW for a node total of $\approx\!8.43$~mW, while BLE's $\sim\!2$~Mbit/s PHY cannot sustain this rate, leaving Wi-R the viable modality. All these totals stay well below full on-device inference, the condition that justifies distribution (Section~\ref{sec:method_energy}).}


\revvv{Hub-side power optimization is outside the scope of this work because DistNN targets the energy-constrained wearable node. For context, Hadidi et al. report average powers of 1.52 W for a Movidius NCS, 4.14 W for an Edge TPU, and 4.58 W for a Jetson Nano, with inference energies from approximately 11 mJ to 1 J~\cite{9041955}. Kim and Wu report peak system powers of 1.8 W with the Hexagon DSP active, 2.8 W with the Adreno GPU, and 5.5 W with the Cortex-A75 CPU, on a Mi8Pro-class mobile SoC~\cite{Kim2020AutoScaleEE}. These values describe complete-network execution on commercial platforms, whereas the DistNN values are node-side estimates for the encoder partition. They therefore provide context rather than a direct comparison.}

\rev{
\paragraph{\textbf{Latency}}
We report node-to-hub latency per inference, $t_{\text{node}} = t_{\text{comp}} + t_{\text{comm}}$, as the time to transmit the payload from node to hub. For AE1, the node runs the encoder in $t_{\text{comp}} = 5.4$~ms at $100$~MHz (obtained from post-synthesis in Section~\ref{ulp}) and outputs a feature map of $DV = 128 \times 8 = 1024$~bits at the distribution point. The communication term $t_{\text{comm}}$ depends on the communication modality. Wi-R streams continuously, so its air time is the payload over the throughput. Using the in-production YR23 chip at $R = 5$~Mb/s~\cite{Ixana81:online} with a conservative efficiency $\eta = 0.5$ for framing and protocol overhead (an effective $R =2.5$~Mb/s), $t_{\text{comm}}^{\text{Wi-R}} = N_{\text{bits}}/(R\eta)$. BLE transmits only during connection events spaced by the connection interval (CI), so its latency is set by how many events the payload needs, $t_{\text{comm}}^{\text{BLE}} = \lceil N_{\text{bits}}/P\rceil\,\text{CI}$. At the minimum CI of $7.5$~ms, one event carries up to $P = 8032$~bits ($1004$~B with packet-length extension on a $2$~Mb/s PHY)~\cite{OnlinePo30:online}. For AE1's $1024$-bit payload, $t_{\text{comm}}^{\text{Wi-R}} = 0.41$~ms and $t_{\text{comm}}^{\text{BLE}} = 7.5$~ms (a single connection event), giving $t_{\text{node}} = 5.8$--$12.9$~ms.
To estimate the end-to-end latency, we add $t_{\text{hub}}$, the decode time at the hub ($\approx\!1.0$~ms, within the $1$--$20$~ms range reported for compact CNNs on mobile GPUs~\cite{lee2019mobilegpu}). For AE1, $t_{\text{e2e}} \approx 6.8$~ms over Wi-R and $\approx 13.9$~ms over BLE, and for AE2, $t_{\text{e2e}} \approx 161$~ms over Wi-R and $\approx 376$~ms over BLE. Table~\ref{tab:latency_strategy} compares the end-to-end per-frame latency of DistNN against running the full network on the node and against sending the raw image to the hub. Note that the full-node time of $103$~ms is the complete AE1 model on the same $100$~MHz cycle-count model as $t_{\text{comp}}$, shown for reference only, since the model does not fit on a wearable node (Section~\ref{sec:rq3_system}).
}\revvv{Against the Category 4 systems of Table III, DDSNN reports 338.7 ms end-to-end and ComAI 536.8 ms per node, both well above DistNN's 6.8–13.9 ms for AE1 and 161 ms for AE2. DISSEC reports 10–25 ms, but on a watt-class Raspberry Pi cluster and without node-power accounting, so the two operate in different power regimes and the latency figures are not directly comparable.}


\begin{table}[t]
  \centering
  \caption{\rev{Per-Frame End-to-End Latency Across Execution Strategies.}}
  \vspace{-2mm}
  \label{tab:latency_strategy}
  \renewcommand\arraystretch{1.3}
  \setlength{\tabcolsep}{3.5pt}
  \scriptsize
  \begin{tabular}{l c c c c c c}
    \hline
     & \multicolumn{2}{c}{\textbf{Full node}} & \multicolumn{2}{c}{\textbf{All hub}} & \multicolumn{2}{c}{\textbf{DistNN}} \\
    \cline{2-3}\cline{4-5}\cline{6-7}
    \textbf{Stage} & \textbf{Wi-R} & \textbf{BLE} & \textbf{Wi-R} & \textbf{BLE} & \textbf{Wi-R} & \textbf{BLE} \\ \hline
    Payload (bits)     & 393{,}216 & 393{,}216 & 393{,}216 & 393{,}216 & 1024 & 1024 \\ \hline
    Node compute (ms)          & 103   & 103   & ---  & ---   & 5.4  & 5.4  \\
    Communication (ms) & 157.3 & 367.5 & 157.3 & 367.5 & 0.41 & 7.5  \\
    Hub (ms)           & ---   & ---   & 1.0  & 1.0   & 1.0  & 1.0  \\ \hline
    \textbf{End-to-end (ms)} & \textbf{260.3} & \textbf{470.5} & \textbf{158.3} & \textbf{368.5} & \textbf{6.8} & \textbf{13.9} \\ \hline 
  \end{tabular}
\vspace{-5mm}
\end{table}

\textbf{\textit{Together, these results place DistNN in a previously underexplored region of the design space. Through this hardware--software co-design, DistNN sustains ULP ($\leq$10mW) node operation while keeping end-to-end latency feasible with low power communication, demonstrating that distributed wearable vision is achievable within ULP budgets.}}



%% file: 6_conclusion.tex
\section{Conclusion}

This work introduced DistNN, a framework for distributed neural-network execution that balances computation between a low-power node and a resource-rich hub. A figure of merit combining local data volume with cumulative computational cost identifies wearable-friendly distribution points from network topology alone, and a custom fixed-point hardware design achieves 230\,fJ/MAC, enabling node-compute power of 0.2 mW (AE1) and 6.5 mW (AE2), or 0.21–8.43 mW including communication. DistNN reaches SSIM values of $\approx$0.90 for image reconstruction tasks while remaining within acceptable wearable power budgets, 
averaging an 11$\times$ reduction in node-compute power relative to the recent ML ASIC baselines evaluated, where the larger gap relative to general-purpose GPU baselines reflects the cost of generality for workloads that do not require it. A current limitation is that the evaluation focuses on vision-oriented autoencoders. However, the underlying principle extends naturally to other architectures because the FoM ranks distribution points from DV and cumulative CC alone, it applies to networks whose feature-map size both contracts, as in convolutional classifiers and residual networks, and expands, as in the autoencoders studied here. This generality, rather than a reliance on monotonic feature-map contraction, is what allows DistNN to support a broader set of sensing and inference tasks. Overall, the framework bridges high computational demand and ultra-low-power operation, providing a path toward energy-efficient wearable inference in resource-constrained environments.
